\documentclass[12pt]{article}
\pdfoutput=1
\usepackage{graphicx,amssymb,amsmath,tensor,empheq,color}

\usepackage{mathtools}

\usepackage{hyperref}
\hypersetup{colorlinks = true, linkcolor=black, citecolor=black, urlcolor=blue}
\usepackage{url}

\makeatletter

\usepackage{esint}

\newlength{\fighskip} \fighskip=2pt
\newlength{\figvskip} \figvskip=3pt

\newcommand*{\figbox}[2]{{
  \def\figscale{#1}
  \def\arraystretch{0.8}
  \arraycolsep=0pt
  \begin{array}{c}
    \vbox{\vskip\figscale\figvskip
      \hbox{\hskip\figscale\fighskip
        \includegraphics[scale=\figscale]{#2}}}
  \end{array}}}
  
\newcommand{\be}{\begin{equation}}
\newcommand{\ee}{\end{equation}}
\newcommand{\bea}{\begin{align}}
\newcommand{\eea}{\end{align}}

\newcommand{\nn}{\nonumber\\}

\newcommand*{\wideboxed}[1]{\setlength{\fboxsep}{1ex}%
  \fbox{\m@th$\displaystyle#1$}}

\def\ubrace#1_#2{%
  \underbrace{#1}_{\hb@xt@\z@{\hss$\scriptstyle#2$\hss}}}

\newcommand\bdot{\mathbin{\mathpalette\bdot@{0.5}}}
\newcommand*\bdot@[2]{\vcenter{\hbox{\scalebox{#2}{$\m@th#1\bullet$}}}}

\newcommand{\hgf}{%
\,\tensor[_{2\kern-1.2pt}]{F}{_{\kern-0.8pt 1}}\kern-1.2pt}
\newcommand{\hgfs}{\mathbf{F}}

\newcommand{\lt}{\left}
\newcommand{\rt}{\right}

\newcommand{\blangle}{\bigl\langle}
\newcommand{\brangle}{\bigr\rangle}
\newcommand{\dlangle}{\langle\kern-1.5pt\langle}
\newcommand{\drangle}{\rangle\kern-1.5pt\rangle}
\newcommand{\bdlangle}{\blangle\kern-3pt\blangle}
\newcommand{\bdrangle}{\brangle\kern-3pt\brangle}

\newcommand*{\bra}[1]{\langle{#1}|}
\newcommand*{\ket}[1]{|{#1}\rangle}
\newcommand*{\braket}[2]{\langle{#1}|{#2}\rangle}

\newcommand*{\bbra}[1]{\blangle{#1}\big|}
\newcommand*{\bket}[1]{\big|{#1}\brangle}
\newcommand*{\bbraket}[2]{\blangle{#1}\big|{#2}\brangle}

\newcommand*{\corr}[1]{\langle{#1}\rangle}
\newcommand*{\bcorr}[1]{\blangle{#1}\brangle}

\newcommand{\vep}{\varepsilon}
\newcommand{\vp}{\varphi}

\newcommand{\calC}{\mathcal{C}}
\newcommand{\calD}{\mathcal{D}}

\newcommand{\calF}{\mathcal{F}}

\newcommand{\calH}{\mathcal{H}}

\newcommand{\calL}{\mathcal{L}}
\newcommand{\calM}{\mathcal{M}}
\newcommand{\calN}{\mathcal{N}}
\newcommand{\calO}{\mathcal{O}}

\newcommand{\calQ}{\mathcal{Q}}

\newcommand{\calU}{\mathcal{U}}

\newcommand{\calX}{\mathcal{X}}

\newcommand{\nab}{\nabla}

\newcommand{\ZZ}{\mathbb{Z}}

\newcommand{\RR}{\mathbb{R}}

\DeclareMathOperator{\tr}{tr}

\DeclareMathOperator{\PSL}{PSL}

\DeclareMathOperator{\tSL}{\widetilde{\mathrm{SL}}}

\DeclareMathOperator{\Sch}{Sch}
\DeclareMathOperator{\ren}{ren}

\DeclareMathOperator{\AdS}{AdS}
\DeclareMathOperator{\tAdS}{\widetilde{AdS}}
\DeclareMathOperator{\HH}{H}

\newcommand{\la}{\text{L}}
\newcommand{\ra}{\text{R}}

\newcommand{\unit}{\mathbf{1}}

\newcommand{\bdy}{\text{bdy}}
\newcommand{\blk}{\text{blk}}

\newcommand{\al}{\alpha}
\newcommand{\bt}{\beta}
\newcommand{\ep}{\epsilon}
\newcommand{\sig}{\sigma}
\newcommand{\tht}{\theta}
\newcommand{\ka}{\kappa}
\newcommand{\ups}{\upsilon}

\newcommand{\Rho}{\mathrm{P}}
\newcommand{\Iota}{\mathrm{I}}

\newcommand{\Sig}{\Sigma}
\newcommand{\lam}{\lambda}

\newcommand{\om}{\omega}

\newcommand{\ga}{\gamma}
\newcommand{\Ga}{\Gamma}
\newcommand{\de}{\delta}
\newcommand{\De}{\Delta}

\def\ie{i.e.\ }
\def\cf{cf.\ }

\newcommand{\ov}{\over}
\newcommand{\p}{\partial}

\makeatother

\title {Dynamics of black holes in Jackiw-Teitelboim gravity}

\author{S.\ Josephine Suh\footnote{suh@caltech.edu}\\
\normalsize\it California Institute of Technology, Pasadena, CA 91125, U.S.A.\vspace{0.5cm}}

\date{May 4, 2020}

\begin{document}

\setcounter{tocdepth}{2}

\maketitle
\thispagestyle{empty}

\begin{abstract}
We present a general solution for correlators of external boundary operators in black hole states of Jackiw-Teitelboim gravity. We use the Hilbert space constructed using the particle-with-spin interpretation of the Jackiw-Teitelboim action, which consists of wavefunctions defined on Lorentzian $\AdS_2$. The density of states of the gravitational system appears in the amplitude for a boundary particle to emit and reabsorb matter. Up to self-interactions of matter, a general correlator can be reduced in an energy basis to a product of amplitudes for interactions and Wilson polynomials mapping between boundary and bulk interactions. 

\end{abstract}

\newpage
\pagenumbering{arabic}
\tableofcontents

\section{Introduction}

Jackiw-Teitelboim (JT) gravity \cite{Ja85, Te83, AlPo14} is the simplest instance of a dilaton gravity in $1+1$ dimensions, in which the dynamical degree of freedom is a boundary or multiple boundaries of a spacetime that has constant negative curvature. Due to its simplicity, it can serve as a laboratory in which to test and further our understanding of various aspects of gravity, in particular those having to do with the emergence of gravity from an underlying microscopic quantum system. This is because the so-called Schwarzian limit of the theory---in which the boundaries are dynamically repelled from self-intersection and have large but finite energy---is recovered as the description at low-temperatures of SYK-like $(0+1)$-dimensional models of holography \cite{SaYe93,Kit.KITP,SoftMode}.

In \cite{KiSuh18}, the present author and Kitaev presented a complete solution to the problem of consistently quantizing JT gravity itself, which has an infinite phase space, without reference to a microscopic system. In particular, we built a Hilbert space of wavefunctions on $\tAdS_2$\footnote{We use this mathematical notation for the universal cover of global $\AdS_2$.} for each boundary of a two-sided black hole and defined on it a finite trace in which the infinite volume of $\tSL(2,\RR)$, the isometry group of $\tAdS_2$ and symmetry of our problem, has been factored out.  As we will review, the starting point for quantizing the theory is to employ an expression for the curvature of a curve valid in two dimensions and which involves the spin connection of the spacetime manifold---the curvature is in fact the Lagrangian of the JT action. Then the action for a black hole boundary is seen to describe a free particle with imaginary spin, and a natural quantization scheme follows in which particle wavefunctions are spinors on $\tAdS_2$ organized into irreducible representations of $\tSL(2,\RR)$.

In order to probe the dynamics of the quantum system thus constructed, one can consider evaluating the correlators of operators inserted on the boundary which emit or absorb matter excitations. In our emergent (from the point of view of a microscopic SYK-like system on the boundary) picture, these matter excitations can be treated as those of quantum fields on $\tAdS_2$. Then a prescription for calculating the matter correlators follows from the Hilbert space of the JT sector described above. Interestingly, it was observed in \cite{KiSuh18} that these correlators satisfy analyticity properties expected of correlators in a micrscopic system, only in the Schwarzian limit and in a combined Hilbert space of gravity plus matter in which, roughly, boundaries propagate in the same time direction as matter fields are quantized.

In this paper, we present a systematic study of correlators of boundary operators in quantum states of the JT black hole, utilizing the symmetry present in Hilbert spaces of both the gravity and matter sectors identified previously. In particular, interactions of boundary and matter, and also between matter, are captured by integrals of wavefunctions over $\tAdS_2$ or asymptotic subspaces of it, and are matrix elements of intertwiners of $\tSL(2,\RR)$. It turns out that as a consequence, correlators---or more generally amplitudes---of boundary operators can be reduced to a simple form by repeatedly applying linear transformations that map a boundary interaction involving both boundary and matter to bulk interactions involving only matter, and vice versa. The same transition amplitude appears in both directions, boundary-bulk and bulk-boundary, and is a Wilson polynomial in boundary energy. 

In the limit of matter fields being free, a general amplitude can be resolved using a linear transformation that maps scattering to non-scattering processes, with the scattering mediated by the boundary. The amplitude that appears in this operation is a gravitational scattering amplitude that can be thought of as being due to t'Hooft shock-waves and which, like the bulk-boundary transition amplitude, originates from a $6j$-symbol involving irreducible representations of $\tSL(2,\RR)$.  Gravitational scattering amplitudes were first discussed in relation to out-of-time order correlators in the Schwarzian theory in \cite{MeTuVe17, Lam18}.\footnote{There the $6j$-symbol appearing in OTOCs  of the Schwarzian theory were obtained by taking the large $c$ limit of a $6j$-symbol of the quantum group $U_q(sl_2)$ inherited from a $2$D CFT.} It turns out that in the presence of bulk interactions they are insufficient for resolving general amplitudes; instead we can use bulk-boundary transition amplitudes---which are more fundamental in the sense that gravitational scattering itself can be decomposed into a series of bulk-boundary ltransitions---in combination with similar operations that reduce bulk interactions.

An outline of this paper as follows: in Section 2, we review the quantization of JT gravity, newly including formulas relevant to the gravitational Hilbert space in the Schwarzian limit. In section 3, we explain our general scheme for resolving amplitudes of boundary operators, and present resulting diagrammatic rules. Main components of the rules include the transition amplitude described thus far, as well as the Schwarzian density of states, which appears in the amplitude for the boundary to emit and reabsorb matter as a nontrivial output of our Hilbert space. In Appendices A and B we provide background for Section 2, and in Appendix C, the calculations using which our diagrammatic rules were derived.

Previous work in which correlators in the Schwarzian theory were analyzed include \cite{MeTuVe17, Bl18, Ya19,Il19}.

\section{Review of quantization of JT gravity}

Here we review the particle formulation of JT gravity and its quantization as a black hole system that was presented in \cite{KiSuh18}, giving some new formulas applicable in the Schwarzian limit.
The gravitational Hilbert space consists of certain irreducible representations of $\tSL(2,\RR)$ realized as spinor wavefunctions on $\tAdS_2$. In the Schwarzian limit, we can consistently enlarge this Hilbert space to include decoupled matter, where wavefunctions resulting from second-quantizing the matter may also be organized into representations of $\tSL(2,\RR)$. The wavefunctions for spacetime and matter in this extended Hilbert space and their symmetry transformations will play an important role in our solution to amplitudes involving the insertion of matter operators on the boundary of spacetime.

\subsection{Particle with imaginary spin}

\begin{figure}[t]
\centerline{\begin{tabular}{c@{\hspace{3cm}}c}
\includegraphics[scale=0.9]{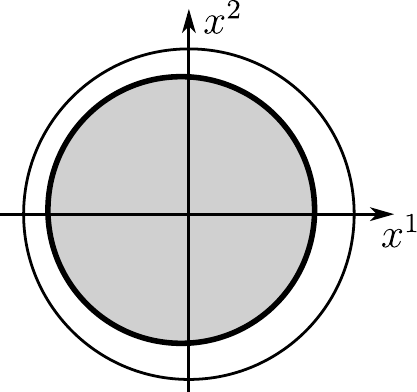} & \includegraphics[scale=0.9]{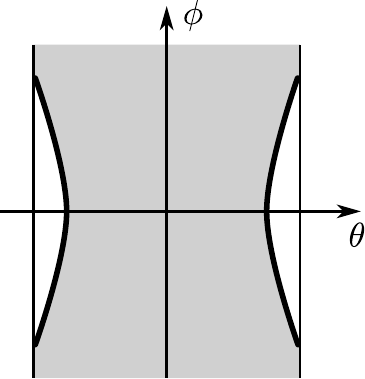}
\vspace{8pt}\\
a) & b)
\end{tabular}}
\caption{a) The regularization of the JT action on $\HH^2$ reveals a free particle with spin. b) We analytically continue the action to $\tAdS_2$, making two copies of the particle, in order to define a two-sided black hole system.  }
\label{fig:geometry}
\end{figure}

The action of JT gravity can be rewritten exactly to reveal a particle degree of freedom with some mass and an imaginary spin. Below, we first explain the appearance of a term in the action which corresponds to particle spin. Then we fix the mass of the particle and regularize the bare action by embedding the system on the hyperbolic disk $\HH^2$---this is because we are interested in studying a two-sided black hole system in JT gravity where each boundary is defined by analytic continuation from $\HH^2$ to $\tAdS_2$, see Figure \ref{fig:geometry}.

Let us consider the action of JT gravity for a two-dimensional spacetime with boundary,
\be\label{IJT}
I_{\text{JT}}[g,\Phi]
=\frac{1}{4\pi}\int_{M}d^2x \sqrt{-g}\, \Phi( R+2)
+\frac{1}{2\pi}\int_{\p M}ds\sqrt{-h}\, \Phi K,
\ee
with the value of the dilaton fixed at the boundary as $\Phi |_{\p M}=\Phi_*$. We may integrate out the dilaton field in the bulk, after which the constraint $R+2=0$ fixes the curvature of spacetime and only the boundary term remains in the action. Then we can view the boundary as a curve or curves embedded in some ambient spacetime of constant negative curvature. Now, simple considerations of the geometry of curves in two dimensions imply that the curvature of a time-like curve $X(u)$ where $u$ is proper time, defined by $\dot{X}^{\nu}\nabla_{\nu}\dot{X}^{\mu}=\kappa n^{\mu}$ where $n$ is the unit normal vector with clock-wise orientation from $\dot{X}$, can be written as
\be \label{kappa}
\kappa=\omega_{\mu}\dot{X}^{\mu}+\dot{\alpha}
\ee
where $\omega_{\mu}=\left(e_1\right)_{\nu}\nabla_{\mu}\left(e_0\right)^{\nu}$ is the spin connection defined using some frame $\{e_0, e_1\}$ on the ambient manifold and $\alpha(u)$ is the angle from $e_0$ to $\dot{X}$. See Figure \ref{fig:kappa}a); details of the derivation are given in Appendix \ref{app:curves}. This curvature in fact agrees with the notion of extrinsic curvature defined for hypersurfaces in general and applied to the curve, $\kappa=-\dot{X}^{\al}\dot{X}^{\bt}\nabla_{\al}n_{\bt}$. Finally, we note that the extrinsic curvature $K$ appearing in \eqref{IJT} should be defined using the unit normal vector that is consistently pointing outward with respect to the spacetime $M$. Thus 
$K=\pm \kappa= \pm \big( \omega_{\mu}\dot{X}^{\mu} + \dot{\alpha}\big)$ depending on whether such a normal vector is parallel or anti-parallel to $n$, see Figure \ref{fig:kappa}b).

\begin{figure}[t]
\centerline{ \begin{tabular}{c@{\hspace{3cm}}c}
\includegraphics[scale=0.9]{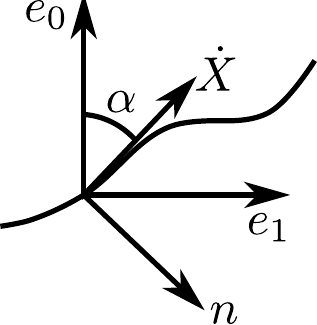} & \includegraphics[scale=0.9]{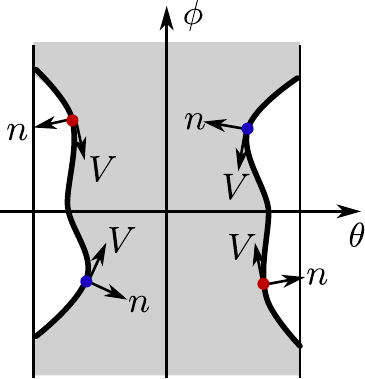}\\
a) & b)
\end{tabular}
}
\caption{a) Definition of angle $\al$ in \eqref{kappa}. b) Red particles have $K=\ka$, and blue particles $K=-\ka$.}
\label{fig:kappa}
\end{figure}

Next, let us put the quantum system defined by \eqref{IJT} on $\HH^2$, that is we fix the ambient spacetime of the fluctuating curve $X$ to be $\HH^2$. A thermal ensemble in this system consists of closed curves that wind around once on the ambient spacetime and have some fixed proper length $L$. We consider separately the cases in which the curves wind clock-wise and anti-clockwise, \ie $\int d\alpha= \pm 2 \pi$. Then fixing proper length with a Lagrange multiplier $E_g$ and also adding a term proportional to $L$ for convenience, the action in a fixed-energy sector of the thermal ensemble is given by (\cf \eqref{IJT})
\begin{align}\label{ptI}
I_{\text{JT bh}}^{\text{Euc}}[X]&=-{\Phi_* \ov 2 \pi}\int ds \sqrt{h}\, (K-1)-E_g \int ds \sqrt{h}\nn
&=\mp \gamma \int dX^{\mu}\om_{\mu}+M \int d\ell- 2 \pi \ga, \qquad \ga={\Phi_* \ov 2 \pi},  \qquad M=\ga-E_g.
\end{align}
In the second line we have used \eqref{kappa}. As referred to earlier, we have arrived at the action for a free particle with mass $M$ and imaginary spin 
\be
\pm\nu, \qquad  \nu=- i \ga.
\ee
In \cite{KiSuh18}, it was shown how to regularize path integrals with the action in \eqref{ptI} by replacing smooth paths with jagged ones consisting of straight segments of a fixed cutoff length. It turns out that the resulting renormalized action is given by
\be  \label{renbhI}
I^{\text{Euc}}_{\text{JT bh,ren}}=\int d\tau\, \left( {	1 \ov 2}g_{\mu\nu}\dot{X}^{\mu}\dot{X}^{\nu}\mp \ga \om_{\mu}\dot{X}^{\mu}\right)
\ee
and that in the Schwarzian limit in which
\be \label{Sch1}
\gamma \gg 1, \qquad L \gg 1
\ee
the renormalized inverse temperature $\beta$ and energy $E$ (conjugate to renormalized proper length $\tau$) are related to their bare counterparts as\footnote{Here $s$ specifies the eigenvalue of the Casimir operator $Q=-L_0^2 +{1 \over 2}\left(L_{-1}L_{1}+L_{1}L_{-1}\right)$ of the Lie algebra $sl_2$ which generates the isometry group of the ambient manifold---$\PSL(2,\RR)$ on $\HH^2$ and $\tSL(2,\RR)$ on $\tAdS_2$---as $q=1/4+s^2$. The parameters $E$ and $s$ are related at the level of quantum equations---for Green functions on $\HH^2$ and for wavefunctions on  $\tAdS_2$---as the Laplacian and $Q$ are related by $-\nabla^2=Q + \nu^2$. \label{Casimir}}

\be \label{ren}
L=\ga\beta, \qquad  \ga E_g=E+{\ga^2 \over 2}-{1 \over 8}={s^2 \ov 2}.
\ee
 We will assign to a single particle corresponding to one boundary of a two-sided black hole in $\tAdS_2$, the action obtained by analytically continuing \eqref{renbhI} back to Lorentzian signature with $\tau=it$,
\be \label{LorI}
I_{\text{JT bh,ren}}=\int dt\, \lt( {1 \ov 2}\,g_{\mu\nu}\dot{X}^{\mu}\dot{X}^{\nu}\pm \ga \om_{\mu}\dot{X}^{\nu}\rt).
\ee
There are two boundaries and thus two particles; one particle has spin $\nu$ and the other $-\nu$, which corresponds to the upper and lower sign in \eqref{LorI}, respectively.



\begin{figure}[t]
\centerline{\begin{tabular}{c@{\hspace{1.5cm}}c}
\includegraphics[scale=0.9]{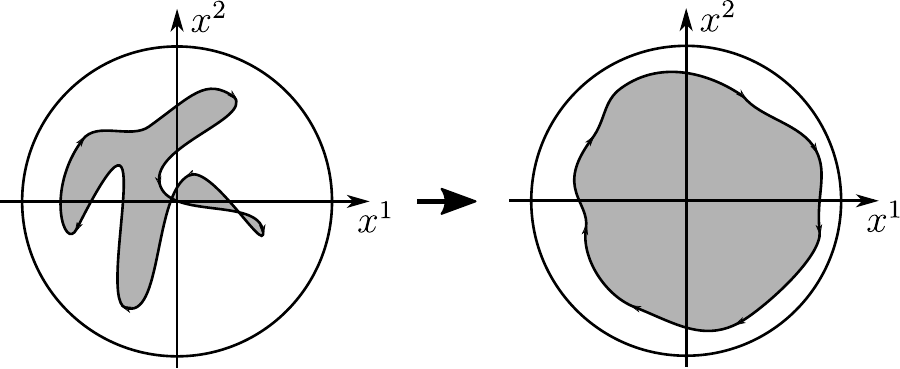}& \includegraphics[scale=0.9]{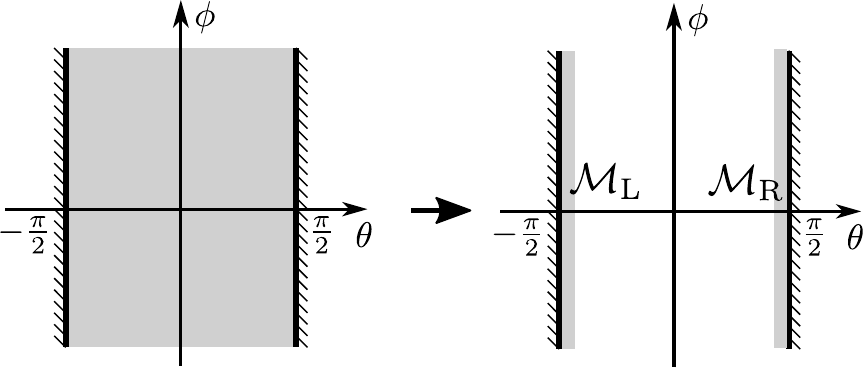}
\vspace{8pt}\\
a) & b)
\end{tabular}}
\caption{a) The Schwarzian limit suppresses self-intersections and pushes the boundary to the asymptotic region of spacetime. b) In $\tAdS_2$, a particle is effectively confined to one of asymptotic spacetimes $\calM_{\rm L}$ or $\calM_{\rm R}$. }
\label{fig:Sch}
\end{figure}

Finally, let us comment on the significance of the Schwarzian limit \eqref{Sch1}, which can also be expressed as 
\be \label{Sch2}
\ga \gg 1, \qquad s^2 \ll \ga^2
\ee
and in which $K-1$ is approximated by a Schwarzian functional of the ambient time coordinate as a function of proper time, see Appendix \ref{app:AdS2_curves}. In this limit a particle trajectory is dynamically induced to be monotonic, \ie its ambient time coordinate as a function of proper time is always positive, and localizes near an asymptotic boundary of the ambient spacetime, see Figure \ref{fig:Sch}. Using parameters that would appear in a microscopic realization of JT gravity by an SYK-like system, we have
\be \label{pmt}
\ga \sim N, \qquad L \sim \beta J , \qquad {\ga \ov L}\sim  {N \ov \bt J}
\ee
Thus the Schwarzian limit corresponds to taking the temperature to be low and $N$ to be large, independently. Meanwhile, the parameter $\ga/L$ which controls quantum effects remains free. We note that it is only in the Schwarzian limit that we can consistently consider second-quantized matter in $\tAdS_2$ together with the black hole system of JT gravity, as we elaborate on in the next section. 


\subsection{Hilbert space}

As mentioned above, our black hole system consists of a $\nu$- and $(-\nu)$-particle in $\tAdS_2$. Here our primary objects of interest will be wavefunctions for a particle resulting from first-quantization.\footnote{These are in fact Wheeler-de Witt wavefunctions for a boundary of spacetime.}. We will use coordinates $(\phi, \theta)$, $-\infty <\phi < \infty$, $-\pi/2 < \tht < \pi/2$ on $\tAdS_2$ in which the metric is
\be \label{AdSg}
ds^2={-d\phi^2 + d\tht^2 \ov \cos^2 \tht}.
\ee
In the Schwarzian limit, particle wavefunctions localize to one of the asymptotic spacetimes $\calM_{\rm R}$ or $\calM_{\rm L}$ on which $\tht \approx {\pi \ov 2}$ and $\tht \approx -{\pi \ov 2}$, respectively. Using the spatial coordinate 
\be \label{asymc}
\upsilon=\ga (\pi \mp 2 \tht)
\ee
they have identical geometry $\calM$ with metric and asymptotic volume form
\be \label{asymg}
ds^2= {-4\ga^2 d\phi^2 +d\upsilon^2 \ov \upsilon^2}, \qquad \int_{\calM}\equiv 2\int_{-\infty}^{\infty} d\phi \int_{0}^{\infty}{d\upsilon \ov \upsilon^2}
\ee

 For purposes of discussion let us fix spin $\nu=-i\ga$. 
 The Schr\"{o}dinger equation derived from \eqref{LorI}, $-{1 \over 2}\nabla^2 \psi=E \psi$, $\nabla_{\mu}=\p_{\mu}+\nu \om_{\mu}$, implies that the wavefunction $\psi$ is a $\nu$-spinor. In explicit terms: $\psi$ is a sectional representation $\psi(x)=\Psi(s(x))$ of a function $\Psi$ defined on $\tSL(2,\RR)$, considered as a principal bundle over $\tAdS_2$ with fiber generated by boost $\Lambda_2$, and which transforms as $\Psi(s(x)e^{-\tht \Lambda_2})=e^{\nu\tht}\psi(x)$. The choice of section $s(x)$ is the gauge for $\psi(x)$; we consistently use the tilde gauge which was defined in \cite{KiSuh18} and in which the frame field is as shown in \eqref{tildeg}. Now, the Schr\"{o}dinger equation is in fact an eigenvalue equation for the Casimir operator of the isometry group $\tSL(2,\RR)$ of $\tAdS_2$ (see Footnote \ref{Casimir}), $Q\psi =\lambda(1-\lambda)\psi$ with $2E=\lambda(1-\lam) -\ga^2.$ Furthermore, common eigenfunctions of $Q$ and $L_0=i \p_{\phi}$, which come in irreducible representations of $\tSL(2,\RR)$, span the space of $\nu$-spinors on $\tAdS_2$ normalizable under the natural inner product 
 \begin{equation} \label{innprod}
\bbraket{\psi_1} {\psi_2}=\int_{\tAdS_2} \psi_1^*(x) \psi_2(x).
\end{equation}

Let us briefly explain the structure of $\tSL(2,\RR)$ irreps that are relevant to our problem of describing a black hole in JT gravity together with matter in $\tAdS_2$. An irrep of $\tSL(2,\RR)$ is characterized by eigenvalues of the Casimir operator and central element of the group, 
\be
Q=\lambda(1-\lam)\quad \text{and} \quad  e^{2\pi i L_0}=e^{-2\pi i \mu}, \,\mu \in \RR/\ZZ.
\ee
In particular, it is spanned by states each with an eigenvalue $L_0=-m$, $m \equiv \mu$ (mod $\ZZ$). As was discussed in \cite{KiSuh18}, the irreps that constitute the Hilbert space for a boundary particle in our problem are those of type
\be \label{prin}
\text{principal series $\calC_{\lambda}^{\mu}$}: \qquad \lambda={1 \over 2}+is,\, s>0, \qquad m=\mu+k ,\,  k \in \ZZ
\ee
with no restriction on the periodicity $\mu$. Note the spin $\pm \nu=\mp i \ga$ of particle wavefunctions have to do with $\textit{how}$ an irrep is realized as a function on $\tAdS_2$, not with specifying the irrep itself. In parallel with particle wavefunctions, we can also consider wavefunctions that result from second-quantizing matter in $\tAdS_2$. These will be realizations of irreps of type
\be \label{dis}
\text{discrete series } \calD^{\pm}_{\lam}: \qquad \lambda>0, \quad \mu=\pm \lambda, \qquad \pm m= \lambda, \lambda+1, \lambda+2,\dots
\ee
with spin that is integer or half-integer. We note that an irrep in the negative (positive) discrete series results from quantizing with respect to time $+\phi$ ($-\phi$), as wavefunctions depend on the $\tAdS_2$ time coordinates $\phi$ as $\psi \sim e^{im \phi}$. This distinction will enter our construction of the total Hilbert space of a black hole plus matter. 

Now, we give an explicit description of the Hilbert space for a $\nu$-particle \textit{after} taking the Schwarzian limit, when it factorizes as $\calH^{ \nu}=\calH^{ \nu}_{\rm R} \otimes \calH^{ \nu}_{\rm L}$ into spaces of wavefunctions localized on $\cal M_{\rm R}$ and $\cal M_{\rm L}$, respectively.
Solving joint eigenvalue equations for $Q$ and $L_0$ in the limit \eqref{Sch2} and in asymptotic regions $\calM_{\rm R, L}$, we find solutions\footnote{Note that taking into account \eqref{prin}, $E=s^2/2+1/8-\ga^2/2$.}
\be \label{psi}
\psi^{\pm}_{\lam, m}(\phi, \upsilon)=\sqrt{{2\pi \rho_{\text{Pl}}(E, \mu) \ov \Ga(\lam \mp m)\Ga(1-\lam \mp m)}}W_{\mp m, is}(\upsilon)\,e^{im \phi}
\ee
which for both signs realize the irrep $\calC^{\mu}_{\lam}$ and respectively span\footnote{The trivially modified wavefunctions $\widetilde{\psi}^{\pm}_{\lam, m}=(-1)^{\mu-m}\psi^{\pm}_{\lam,m}$ span $\widetilde{\calH}^{+}=\calH^{-\nu}_{\rm L}$ and $\widetilde{\calH}^{-}=\calH^{-\nu}_{\rm R}$; hence the notation $\pm$ which refers to the direction of time, see \eqref{tHS} and following explanation. \label{ospin}}
\be
\calH^{+}=\calH_{\rm R}^{\nu}, \qquad \calH^{-}=\calH_{\rm L}^{\nu}.
\ee
In \eqref{psi} $W_{\al, \bt}(z)$ is the Whittaker hypergeometric function, and in the normalization we have used the Plancherel measure for $\tSL(2,\RR)$ irreps in the principal series,
\be \label{Plm}
\rho_{\text{Pl}}(E,\mu)=(2\pi)^{-2} {\sinh (2 \pi s) \ov 2 b}, \qquad b={1 \over 2}\left( \cosh(2\pi s)+\cos (2\pi \mu)\right)
\ee
which is an effective dimension of an $\tSL(2,\RR)$ irrep with the infinite volume of the group factored out. The normalization has been chosen such that (\cf \eqref{innprod})
\begin{align}
\bcorr{\psi_{1/2+ is, m}^{\pm} | \psi^{\pm}_{1/2+is', m'}}_{\text{Sch}}&=\int_{\calM} \left(\psi^{\pm}_{1/2+is, m}\right)^*\psi^{\pm}_{1/2+i s', m'}\nn
&=\de(E-E')\de(m-m').
\end{align}
Then a general $\tSL(2,\RR)$-invariant operator on $\calH^{\pm}$ takes the form
\be \label{op}
\Psi^{\pm}[f]=\int dE \underbrace{\int d\mu\, f(E, \mu) \sum_{m \in \mu + \mathbb{Z}} \bket{\psi^{\pm}_{\lambda, m}}\bbra{ \psi^{\pm}_{\lambda, m}}}_{\equiv \left(\Psi^{\pm}[f]\right)_E}
\ee
where we have access to a state vector $\ket{\psi}$ through its wavefunction $\braket{x  }{\psi}=\psi(x)$, and the trace over $\calH^{\pm}$ of a product of such operators is given by\footnote{Note the absence of the overall factor of $1/2$ that was present in the trace defined in \cite{KiSuh18}, which was over a two-sided Hilbert space.}
\be \label{trdef}
\text{tr} (\Psi^{\pm}[f_1]\dots\Psi^{\pm}[f_n])=\int dE \int d\mu\, \rho_{\text{Pl}}(E, \mu) f_1(E, \mu)\dots f_n (E, \mu).
\ee

\begin{figure}[t]
\centerline{\begin{tabular}{c@{\hspace{3cm}}c}
\includegraphics{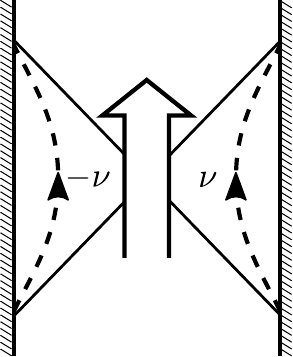} & \includegraphics{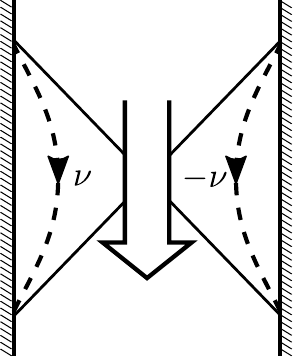}  \\
a) & b) 
\end{tabular}}
\caption{a,b) The time with respect to which matter fields should be quantized, determined by the direction of propagation for $\nu$- and $(-\nu)$-particles on opposite sides}
\label{fig:time}
\end{figure}

Finally, we consider adding matter fields to the black hole system in $\tAdS_2$. In \cite{KiSuh18} it was shown that the total Hilbert space of the black hole plus matter without explicit couplings to the Schwarzian sector, should take the form
\begin{equation}\label{tHS}
\calH = \bigl(\calH_{\text{fields}}^{+}
\otimes\calH^{\nu}_{\ra}\otimes\calH^{-\nu}_{\la}\bigr)
\oplus\bigl(\calH_{\text{fields}}^{-}
\otimes\calH^{\nu}_{\la}\otimes\calH^{-\nu}_{\ra}\bigr)
\end{equation}
where $\calH_{\text{fields}}^{\pm}$ denotes a Hilbert space of fields which have been quantized with respect to time $\pm\phi$, which is the same direction as boundary particles propagate. See Figure \ref{fig:time}. The consistency of our general solution for arbitrary correlators of operators on the boundary will depend on using the Hilbert space in \eqref{tHS}. The two sectors in the direct sum are decoupled; then for example an operator acting in the first sector and on the $\nu$-particle will take the form
\be\label{O_R}
\hat{\calO}
=\int_{\calM_{\rm R}} \calO(x)\otimes \ket{x}\bra{x}\otimes \unit,\qquad
\hat{\calO}(T)=e^{iHT}\hat{\calO}e^{-iH T}
\ee
where $H=\Psi^{+}[E]$. Note that a boundary operator constructed as above is directly an operator in the bulk field theory, as opposed to being an operator in a boundary theory dual to a bulk field. In order to obtain cutoff or $\ga$-independent correlators, it is necessary to use renormalized boundary operators
\be \label{renorm}
\hat{\calO}\, \to \,\hat{\calO}_{\ren}=\ga^\De \hat{\calO}, \qquad \De=\text{dim}[\calO].
\ee
This is akin to holographic renormalization of bulk fields dual to boundary operators in the AdS/CFT setting, see \eqref{pmt}. From here on we will always assume that boundary operators $\hat{\calO}$ are implicitly renormalized.

In constructing density matrices on $\calH^{\pm}$, it was shown in \cite{KiSuh18} that the continuous density of states 
\be \label{rho}
\rho_{\text{Sch}}(E)=\pi^{-2}\sinh 2 \pi s
\ee
is accounted for by an operator 
\be \label{P}
{\rm P}=\Psi^{\pm}[8 b], \quad  \text{tr}({\rm P})=\int dE\, \rho_{\text{Sch}}(E)
\ee
which captures the amplitude for the boundary particle to tunnel to the opposite side then back. So for instance the thermal density matrix of the boundary is given by $\varrho_{\bt}=Z^{-1}e^{-\bt H}\, {\rm P}$, $Z=\int dE\, e^{-\bt E} \rho_{\text{Sch}}(E)$, and the two-point function of operator $\hat{\calO}$ in the thermal state, by
\be \label{twopcorr}
\left\langle\text{tr}\left( Z^{-1}e^{-\bt H} {\rm P}\, \hat{\calO}(T)\hat{\calO}(0)\right)\right \rangle
\ee
where the outer expectation value indicates the evaluation in some field theory state in $\calH^{\pm}_{\text{fields}}$, of the operator product $\calO(x)\calO(x')$ appearing in $\hat{\calO}(T) \hat{\calO}(0)$. In our general evaluation of amplitudes, it will turn out that $\rho_{\text{Sch}}(E)$ in \eqref{rho} appears automatically in the kernel of the operator $\corr{{\rm I}_{E''}\hat{\calO} {\rm I}_{E} \hat{\calO}{\rm I}_{E'}}$ where
\be \label{proj}
{\rm I}=\Psi^{\pm}[1]=\int dE \int d\mu\,  \sum_{m \in \mu + \mathbb{Z}} \bket{\psi^{\pm}_{\lambda, m}}\bbra{ \psi^{\pm}_{\lambda, m}}
\ee
is the particle propagator---this kernel is the amplitude in energy basis for the boundary particle to emit and reabsorb a matter excitation. Neither wavefunctions $\psi^{\pm}_{\lam, m}$ nor the propagator ${\rm I}_E$ know about $\rho_{\text{Sch}}$, only $\rho_{\text{Pl}}$, so this appearance is an intrinsically Lorentzian derivation of the density of states of the gravitational system.\footnote{In comparison, the operator ${\rm P}$ was essentially obtained by squaring the analytic continuation of the particle two-point function on $\HH^2$.}

For purposes of calculating amplitudes in the next section, let us specify the matter sector. To be concrete, we will assume fields in the bulk are weakly interacting scalars, and the field theory is in its vacuum state. Then the propagator for a field with dimension $\De>1/2$ is given by 
\be \label{twop}
\corr{\calO(x)\calO(x')}=\sum_{k=0}^{\infty}\psi^{0}_{\De,\mp(\De+k)}(x)\,\psi^{0}_{\De,\mp(\De+k)}(x')^{*}
\ee
where $\big\{\bket{\psi^{0}_{\lam,m=\mp(\lam+k)}}\big\}$ is a realization of the discrete irrep $\calD^{\mp}_{\lam}$ by spin-$0$ wavefunctions on $\tAdS_2$. The overall normalization of these wavefunctions is in principle arbitrary, with corresponding factors that will simply multiply amplitudes in the theory. We fix it so that using the inner product in \eqref{innprod}, $\bcorr{\psi^{0}_{\De, \mp(\De+k)} | \psi^{0}_{\De',\mp(\De'+k')}}=\de(\De-\De')\de_{k, k'}$. 
Explicit expressions are given in Appendix \ref{app:scwf}. 

\section{Evaluation of correlators}

\begin{figure}[t]
\centerline{
\begin{tabular}{c@{\hspace{3cm}}c@{\hspace{3cm}}c}
\includegraphics[scale=1.3]{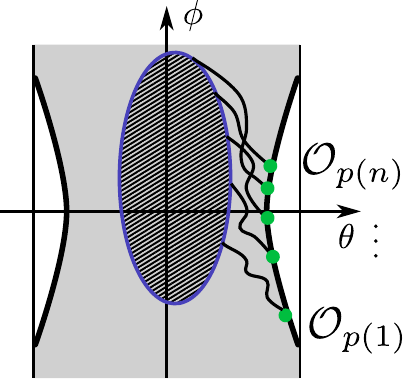} & \includegraphics[scale=1.5]{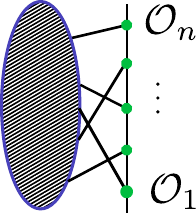} & \includegraphics[scale=1.5]{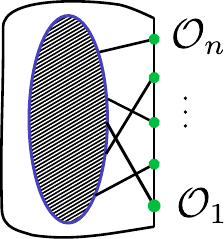}\\
&\vspace{1pt}  &\\
a) & b) & c)
\end{tabular}
}
\caption{a) A spacetime depiction of a correlator of boundary operators. Representations in Hilbert space of b) an amplitude and c) a correlator. In general, the time-ordering of operators is different from their ordering in Hilbert space.}
\label{fig:corr}
\end{figure}

We are now ready to present a general analysis of correlators of matter operators in black hole states of the boundary. We will actually analyze amplitudes for operators to act on a boundary, which are kernels of operators of the form 
\be\label{ampop}
\corr{{\rm I}\hat{\calO}_{n}\dots {\rm I} \hat{\calO}_{1}{\rm I}}
\ee
Using only wavefunctions presented in the previous section as input, we will arrive at a set of diagrammatic rules, which together with rules to resolve possible interactions of bulk fields, may be used to evaluate an arbitrary amplitude. When applied to diagrams with closed boundary, these rules will reproduce black hole expectation values computed using the trace in \eqref{trdef} and tunneling operator squared in \eqref{P}, $\bcorr{\tr({\rm P} \,\hat{\calO}_{n}\dots \hat{\calO}_{1})}$. See Figure \ref{fig:corr}.

Modulo the density of states which appears whenever integrating over energies of boundary propagators, there will be two main components to the diagrammatic rules: i) amplitudes associated with interactions which can be either on the boundary or in the bulk, and ii) transition amplitudes between boundary and bulk interactions. A general amplitude will be reduced to a product of these amplitudes in some energy basis. We also identify amplitudes for gravitational scattering of matter in our black hole system (via shock waves), which if the matter fields are free can be used for reduction of diagrams in place of boundary-bulk transition amplitudes. 

Throughout this section, we work in the Hilbert space \eqref{tHS}. In particular, we consider particle wavefunctions in $\calH^{\pm}$ (or equivalently $\widetilde{\calH}^{\pm}$, see Footnote \ref{ospin})  together with matter wavefunctions in $\calH_{\text{fields}}^{\pm}\cong \oplus_{\De}\calD^{\mp}_{\De}$, with signs correlated. We will suppress signs differentiating between the two cases, except when they are necessary to specify wavefunctions or representations in $\calH_{\text{fields}}^{\pm}$, or otherwise enter the discussion explicitly. It is to be assumed that integrals involving wavefunctions in $\calH^{+}$ ($\widetilde{\calH}^-$) are over $\calM_{\rm R}$, and those involving wavefunctions in $\calH^{-}$ ($\widetilde{\calH}^+$), over $\calM_{\rm L}$. 

\subsection{General scheme} \label{sec:scheme}

The main idea of our analysis is to break up propagators \eqref{proj} and \eqref{twop} appearing in \eqref{ampop} in half, and to decompose the kernel of the operator (in the notation of \eqref{op} we refer to $f$ as the kernel of the operator $\Psi[f]$) in terms of integrals of wavefunctions over $\calM$ occurring at each operator insertion. (Interactions involving matter wavefunctions only will produce integrals over $\tAdS_2$.) By construction, these integrals define intertwiners that commute with the action of $\tSL(2,\RR)$ on the wavefunctions.

\begin{figure}[t]
\centerline{
\includegraphics[scale=1]{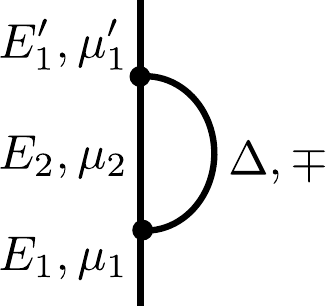}
}

\caption{Diagram corresponding to $\ker\big[\bcorr{\Iota_{E'_1}\hat{\calO}_{\De}\Iota_{E_2}\hat{\calO}_{\De} \Iota_{E_1}}\big]$. }
\label{fig:reabsorb}
\end{figure}

For illustration and to fix diagrammatic notation, let us consider the emission and reabsorption of a matter excitation of dimension $\De$ by the boundary. Using \eqref{O_R}, \eqref{renorm}, \eqref{proj}, and \eqref{twop}, 
\begin{align}
\bcorr{\Iota_{E_1'}\hat{\calO}_{\De}\Iota_{E_2}\hat{\calO}_{\De} \Iota_{E_1}}&=\int d\mu_1 \sum_{m_1\in \mu_1 +\ZZ}\int d\mu_2\sum_{m_2 \in \mu_2+\ZZ}\int d\mu'_1 \sum_{m'_1 \in \mu'_1+\ZZ}\ket{\psi_{\lam'_1, m'_1}}\bra{\psi_{\lam_1,m_1}}\nn
&\underbrace{\sum_{k=0}^{\infty}\int_{\calM}\left(\psi_{\lam'_1, m'_1}\right)^*\cdot \psi_{\lam_2, m_2}\cdot\ga^{\De}\psi^{0}_{\De,\mp( \De+k)}\int_{\calM}\left(\psi_{\lam_2, m_2}\right)^*\cdot\ga^{\De}\left(\psi^{0}_{\De, \mp(\De+k)}\right)^*\cdot \psi_{\lam_1,m_1} }_{\propto\, \de(m'_1-m_2\pm (\De+k))\de(m_2\mp(\De+k)-m_1)\de(E'_1-E_1)} \nn
&=\int d\mu_1\, \ker\big[\bcorr{\Iota_{E'_1}\hat{\calO}_{\De}\Iota_{E_2}\hat{\calO}_{\De} \Iota_{E_1}}\big]\sum_{m_1 \in \mu_1+\ZZ} \ket{\psi_{\lam_1, m_1}}\bra{\psi_{\lam_1,m_1}}.
\end{align}
That the expression in the second line is proportional to denoted delta functions is due to the intertwiner property of the integrals; in addition it is guaranteed to be independent of $m$, which leads to the operator expression in the last line. We denote the kernel of this operator by the diagram in Figure \ref{fig:reabsorb}. It is to be understood that $\mu$-parameters are conserved at each vertex, for example $\mu_1=\mu_2 \mp \Delta$ (mod $1$). 

More generally, any diagram involving interactions between boundary and matter (``boundary") or just matter (``bulk") may be viewed as a linear map from the tensor product of $\tSL(2,\RR)$ irreps to which legs on the bottom  (``in states") belong, to the tensor product of $\tSL(2,\RR)$ irreps to which legs on the top (``out states") belong---where the irreps are realized by wavefunctions given in the previous section. It follows that a general amplitude can be solved for by repeated application of changes of bases of maps that resolve a complicated unit of diagram into a linear combination of simpler diagrams that have the same in and out states. In our problem, it will be sufficient to resolve certain unit diagrams involving four irreps; in representation-theoretic language, the coefficients appearing in such changes of bases are $6j$-symbols. 

We also note that a priori, an arbitrary diagram may depend on the $\mu$-parameters of its legs, and also whether we work in the first or second factors of \eqref{tHS}, However, it will turn out that amplitudes of our interest, \ie kernels of operators of the form \eqref{ampop}, will be independent of these considerations. Therefore in what follows we will abbreviate $\mu$-parameters and $\mp$ labels on legs and propagators of all diagrams.

\begin{figure}[t]
\centerline{\begin{tabular}{c @{\hspace{4.5cm}}c}
\includegraphics[scale=0.8]{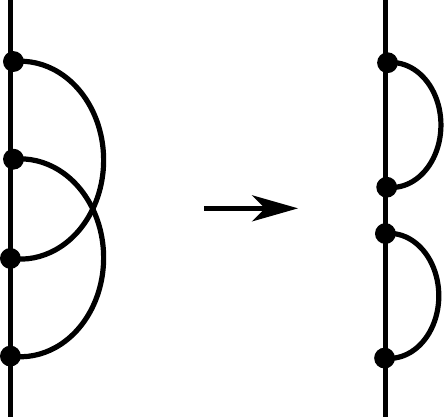}& \includegraphics[scale=0.6]{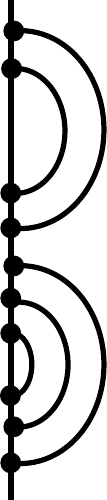}\\
a) & b)
\end{tabular}
}
\caption{a) Reduction of an amplitude by uncrossing. b) A maximally reduced amplitude.}
\label{fig:uncrossed}
\end{figure}

Let us first consider the limit of matter fields being free. Then to resolve any amplitude it is sufficient to use a change of basis of maps which we call an uncrosser and define as
\be \label{uncross}
\figbox{0.8}{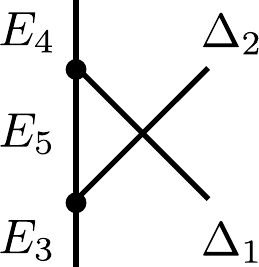}\quad = \int dE_6 \, F^{E_4, \De_2}_{E_3, \De_1}(E_5; E_6)\quad \figbox{0.8}{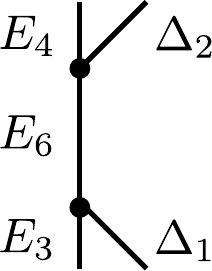} 
\ee
After repeated uncrossings, an amplitude can be reduced to a form that can be evaluated using only emission-reabosorption amplitudes, see Figure \ref{fig:uncrossed}. The physical significance of the coefficients of uncrossing can be seen in Figure \ref{fig:uncrossed}a). Before uncrossing, the diagram computes an out-of-time order 4-point amplitude, and after, a time-ordered 4-point amplitude; therefore the coefficients of uncrossing appear in the ratio of a 4-point OTOC to a 4-point TOC, and should contain the amplitude for gravitational scattering via shock waves. In the next section, we will isolate in a precise manner the gravitational scattering amplitude.

\begin{figure}[t]
\centerline{\begin{tabular}{c}
\includegraphics[scale=0.8]{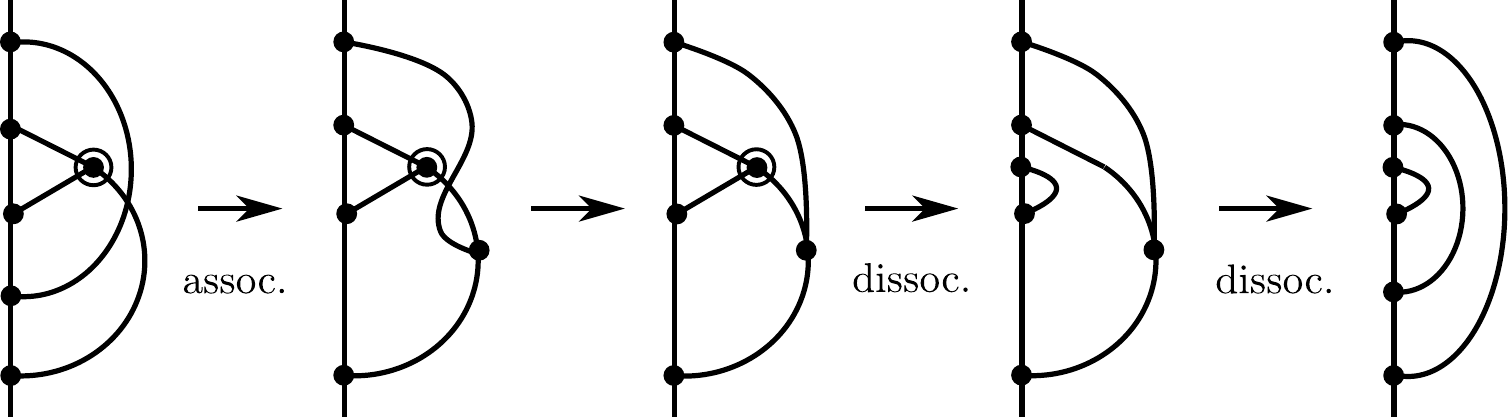}
\end{tabular}
}
\caption{Reduction of an amplitude involving a bulk interaction (ringed vertex) using associators and dissociators. In the second step we exchange two discrete irreps at a bulk interaction which gives a minus sign.}
\label{fig:resolve}
\end{figure}

When interactions of matter fields are turned on, uncrossings are inadequate for resolving general amplitudes. Instead, we can use the pair of changes of bases
\begin{align} 
\label{ass}
\figbox{0.75}{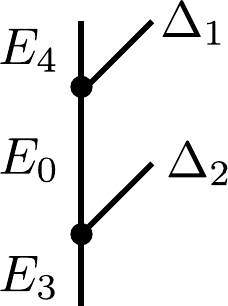}\quad&=\sum_{n=0}^{\infty}F^{E_4, \De_1, \De_2}_{E_3}(E_0; n)\quad\figbox{0.8}{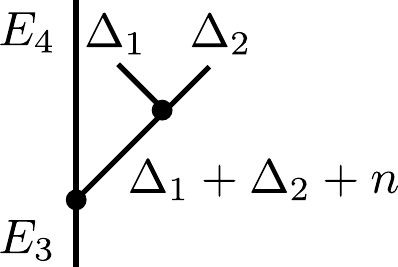}\\
\label{diss}
\figbox{0.8}{assoc_wide}\quad&=\int dE_0\, \left(F^{E_4, \De_1, \De_2}_{E_3}\right)^{\dagger}(n; E_0)\quad\figbox{0.75}{unassoc}
\end{align}
which we call an associator and dissociator, respectively.\footnote{The conjugation in \eqref{diss} is respect to the natural Hermitian inner product on the spaces of maps $\calH_{\lam_3}^{\mu_3} \to \calH^{\mu_4}_{\lam_4} \otimes \left(\calH_{\text{fields}}\right)^{\mp}_{\De_1}\otimes\left(\calH_{\text{fields}}\right)^{\mp}_{\De_2} $, analogous to the inner product demonstrated in \eqref{innprodapp} of Appendix \ref{app:6j}.} 
In combination with analogous operations involving only irreps in $\calH_{\text{fields}}$ that can be used to simplify bulk interactions, they are sufficient for reducing a general diagram. See Figure \ref{fig:resolve} for an example of their use. The coefficients of an associator or dissociator contain a transition amplitude between boundary and bulk interactions, which we will also isolate in the next section.

Finally, let us describe how we obtain $6j$-symbols appearing in the above changes of bases, as well as basic diagrams required for evaluating completely reduced amplitudes. We make use of the fact that an $\tSL(2,\RR)$ irrep labeled by $\lambda$ and $\mu$ can be embedded into the space of $\mu$-twisted $\lambda$-forms on a circle, which are functions $f(\vp)$ obeying $f(\vp+2\pi)=e^{2\pi i\mu }f(\vp)$ and $\left(V f\right)(\varphi)=\left(\p_{\vp}V^{-1}(\vp)\right)^{\lam}f(V^{-1}(\vp))$ for a diffeomorphism of the circle $V$. There are two embeddings
\be \label{embed}
\Xi^{\mu \pm}_{\lam}: \calU^{\mu}_{\lambda} \to  \calF^{\mu}_{\lam},\qquad  \Xi^{\mu \pm}_{\lam}\ket{m}=c^{\pm}_{\lam, m}f_{\lam, m}
\ee
where $f_{\lam, m}=e^{im\vp}$ and we give the coefficients $c^{\pm}$ in Appendix \ref{app:calc}. If $\calU^{\mu}_{\lam}$ is a discrete irrep $\calD^{\pm}_{\lam}$, only embeddings with corresponding signs exist; see \cite{SL2R} for further facts about these embeddings. Given a diagram in our theory, it is simpler to first evaluate it interpreting each vertex as the kernel of a map involving spaces $\calF^{\mu}_{\lam}$ rather than the physical Hilbert spaces in our problem. Such a kernel, up to coefficients of embeddings in \eqref{embed}, in turn defines by commutation a map involving spaces $\calU^{\mu}_{\lam}$. See the figure in \eqref{commapp} for an example. In order to recover the original diagram involving Hilbert spaces of wavefunctions, we only have to multiply at each (three-way) vertex the ratio of the integral of wavefunctions defining the original diagram over the kernel of the map involving $\calU_{\lam}^{\mu}$. The latter are Clebsch-Gordon coefficients and the ratio is some overall factor that does not depend on the $m$-labels of in and out states as our physical Hilbert spaces are in fact isomorphic to $\tSL(2,\RR)$ irreps. We carry out these steps explicitly in Appendix \ref{app:calc}.

\subsection{Diagrammatic rules}

After evaluating amplitudes for emission-reabsorption and splitting-remerging of matter excitations, as well as $6j$-symbols in the relations \eqref{uncross}, \eqref{ass}, and \eqref{diss}, we can identify the following distinct components: the density of states in the Schwarzian and matter sectors $\rho_{\Sch}(E)$, $\rho_m(\De)$,\footnote{$\rho_m$ is the density of states in the matter sector in the sense that it would be the measure if we were to integrate over the dimensions of scalars $\Delta$ on a propagator. In our actual problem, we do not integrate over $\Delta$ so $\rho_m$ is merely a factor associated with each propagator resulting from the norm that we have chosen for matter wavefunctions.} amplitudes associated with boundary and bulk interactions $A_{\bdy} (E_1, E_2, \Delta)$, $A_{\blk}(\De_1, \De_2, n)$, and amplitudes for gravitational scattering and transition between boundary and bulk interactions, $S^{E_4, \De_2}_{E_3, \De_1}(E_5; E_6)$ and $T_{E_3}^{E_4, \De_1, \De_2}(E_5; n)$, respectively. These designations of components are justified by their appearance in the following:
\begin{align} 
\label{diag1}
&\quad\quad\figbox{0.8}{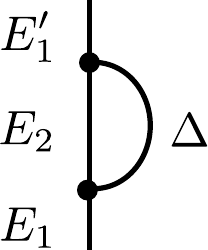}\quad =\de(E_1-E'_1)\rho_{\Sch}(E_2)\rho_m(\De)A_{\bdy}(E_1, E_2, \De),\\[10pt]
\label{diag2}
&\figbox{0.8}{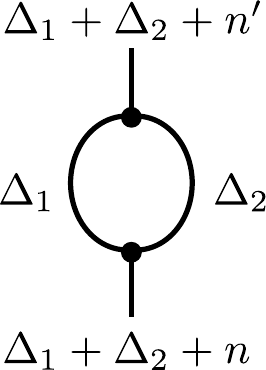}\quad=\de_{n,n'}\rho_m(\De_1)\rho_m(\De_2)A_{\blk}(\De_1, \De_2, n),\\[10pt]
\label{diag3}
&\figbox{0.8}{cross} =\rho_{\Sch}(E_5)A_{\bdy}(E_3, E_5, \De_2)A_{\bdy}(E_5, E_4, \De_1)\int d E_6\,  S^{E_4, \De_2}_{E_3, \De_1}(E_5; E_6)\, \,\figbox{0.8}{uncross},
\end{align}
and with $\De=\De_1+\De_2+n$,
\begin{align} 
\label{diag4}
&\figbox{0.75}{unassoc}  \,=\rho_{\Sch}(E_0)A_{\bdy}(E_3, E_0, \De_1)A_{\bdy}(E_0, E_4, \De_2) \sum_{\substack{n=2l\\ l=0,1,\dots }}T_{E_3}^{E_4, \De_1, \De_2}(E_0; n)\figbox{0.8}{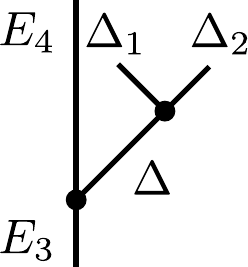},\\[5pt]
\label{diag5}
&\figbox{0.8}{assoc}  \,=\rho_{m}(\De)A_{\bdy}(E_3, E_4, \De)A_{\blk}(\De_1, \De_2, n) \int dE_5\, T_{E_3}^{E_4, \De_1, \De_2}(E_5; n)\,\figbox{0.75}{unassoc}.
\end{align}

The density of states in the Schwarzian black hole system $\rho_{\Sch}$ coincides with the expression that was given in \eqref{rho}; as noted previously, its appearance in the calculations above \eqref{diag1}-\eqref{diag5} is a non-trivial output of the wavefunctions \eqref{psi} in our gravitational Hilbert space, where the latter followed straightforwardly from the Lorentzian action \eqref{LorI}. The matter density of states and operator amplitudes are given by
\be \label{rhom}
\rho_{m}={\Ga(\De)^2 (\De-{1 \ov 2}) \ov (2\pi)^2 \Ga(2 \De)},
\ee
\begin{align} \label{amp}
&A_{\bdy}(E_1, E_2, \De) =\Ga(\De\pm i s_1 \pm is _2),\\
&A_{\blk}(\De_1, \De_2, n=2l)=2\pi\left(\cos{\pi n \over 2}\right)^2 {\Gamma(l+{1 \over 2}) \over \Gamma(l+1)}2 ^{2(\Delta_1+\Delta_2)}{\Gamma(\Delta_1+\Delta_2 + l-{1 \over 2}) \left(\Delta_1\right)_l (\Delta_2)_l\over \Gamma(\Delta_1+\Delta_2+l)\left(\Delta_1+{1 \over 2}\right)_l \left(\Delta_2+{1 \over 2}\right)_l},
\end{align}
where we have used the notation that a product is taken over alternating signs inside a gamma function. The gravitational scattering amplitude is
\be \label{Samp}
S^{E_4, \De_2}_{E_3, \De_1}(E_5; E_6)=\phi_{s_3}\left( s_4; \lambda_6+\Delta_2-{1 \over 2},  \Delta_1+\lambda_5-{1 \over 2},  \Delta_1-\lambda_5+{1 \over 2},  \lambda_6-\Delta_2+{1 \over 2}\right)
\ee
where 
\begin{align} \label{Wilf}
\phi_{\lambda}\left( x; a, b,c, d\right)&={\Gamma(\tilde{a}+\tilde{b}+\tilde{c}+i \lambda) \over \Gamma(a+b)\Gamma(a+c)\Gamma(1+a-d)\Gamma(1-\tilde{d}-i \lambda)\Gamma(b+c+i \lambda\pm i x)}\nn
&\times W\left( \tilde{a}+\tilde{b}+\tilde{c}-1+i \lambda; a+i x, a-i x, \tilde{a}+i \lambda, \tilde{b}+i \lambda, \tilde{c}+i \lambda\right)
\end{align}
is the Wilson function defined in (4.4) of \cite{Gro05}, proportional to a well-poised $_7 F_6$ hypergeometric function (sum of balanced $_4 F_3$ hypergeometric functions) of unit argument. The transition amplitude between boundary and bulk interactions is given by
\begin{align} \label{Tamp}
&T_{E_3}^{E_4, \Delta_1, \Delta_2}\left(E_0; n\right)=\\
&{1 \over \Gamma(\Delta\pm is_3  \pm is_4)}{(-1)^l \Gamma(l) \over 2 \Gamma(2 l)}{\Gamma(\Delta_1+\Delta_2+ l) \over \Gamma(\Delta_1+l)\Gamma(\Delta_2+l)}{\Gamma(\Delta_1)\Gamma(\Delta_2) \over \Gamma(\Delta)}\sqrt{{\Gamma(2 \Delta) \over \Gamma(2 \Delta_1)\Gamma(2 \Delta_2)}} \\
&\times W_{n}(s_0; \Delta_1+i s_3, \Delta_1-i s_3, \Delta_2+i s_4, \Delta_2-i s_4)
\end{align}
where $\De=\De_1+\De_2+n$ for $n=2l$, $l \in \ZZ^{+}$, and zero otherwise, and
\begin{align} \label{Wilp}
W_n(x; a, b, c, d)&=(a+b)_n(a+c)_n(a+d)_{n} \nn
&\times {}_{4} F_3\left( -n, n+a+b+c+d-1, a+i x, a-i x; a+b, a+c, a+d; 1\right)
\end{align}
is the Wilson polynomial of order $n$ in $x^2$, symmetric in $a, b, c, d$. These polynomials satisfy orthogonality relations studied in \cite{Wil80} and have $q$-analogs, the Askey-Wilson polynomials. 

We note that there are no dependences on $\mu$-parameters of irreps of the boundary Hilbert space in the above results. This is a necessary condition for consistency with the analytic continuation of amplitudes from the hyperbolic disk, and in fact arises as a non-trivial consequence of working in the Hilbert space \eqref{tHS}. See Appendix \ref{app:calc} for more details.

\begin{figure}[t]
\centerline{\begin{tabular}{c}
\includegraphics[scale=0.77]{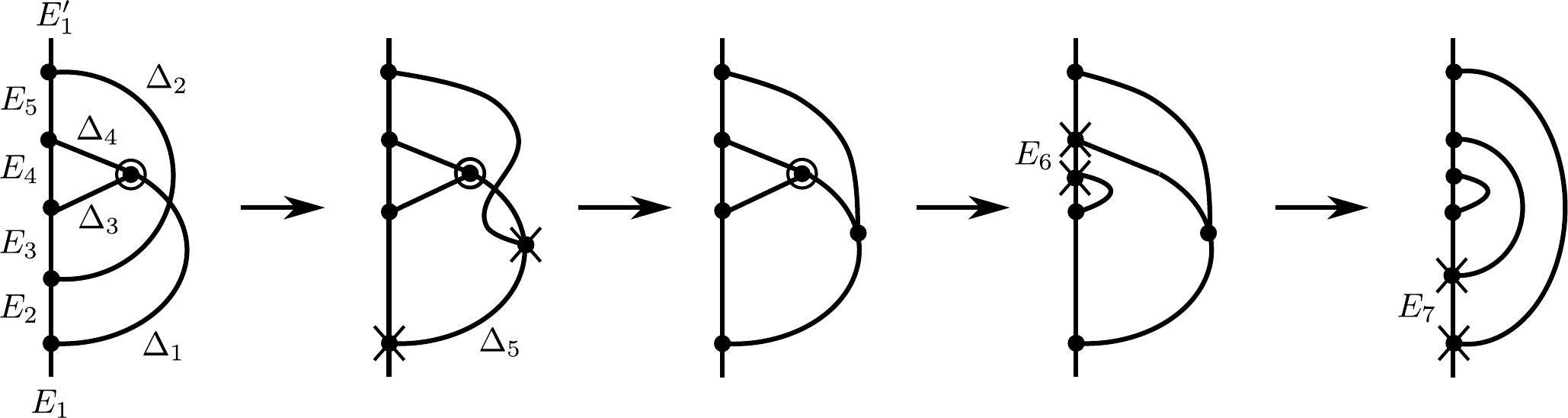}
\end{tabular}
}
\caption{An application of our diagrammatic rules involving the reduction steps shown in Figure \ref{fig:resolve}. At each step, new vertices are marked with a cross and only new propagators are labeled.}
\label{fig:resolve_rules}
\end{figure}

Finally, we present diagrammatic rules for evaluating amplitudes which are established by induction using \eqref{diag1}-\eqref{diag5}. Given a diagram, one should:

\begin{enumerate}

\item
Assign distinct energies to boundary propagators and legs, with the caveat that energy is conserved after consecutive emission and reabsorption of a matter excitation. Similarly, the dimension of a matter propagator is conserved after consecutive splitting and merging.

\item
Multiply by $\rho_{\Sch}$ for each boundary propagator and $\rho_{m}$ for each bulk propagator.

\item
Multiply by $A_{\bdy}$ and $A_{\blk}$ for \textit{distinct} boundary and bulk vertices, respectively. (For example, in \eqref{diag1}, there is only one factor of $A_{\bdy}$.) In the case that a bulk vertex denotes a dynamic interaction rather than an interaction induced by reduction, multiply by dynamically determined amplitude.

\item
Undo gravitational scattering of matter excitations by multiplying scattering amplitudes $S$. More generally, reduce the diagram by moving interactions from boundary to bulk or vice versa by multiplying transition amplitudes $T$, while simultaneously simplifying bulk portions of the diagram as necessary.

\item
Repeat the above sequence of steps, applying steps $1$-$3$ only to new propagators and vertices having resulted from step $4$, until the diagram reduces to a simple product of amplitudes $A_{\bdy}$ and $A_{\blk}$.

\item
Integrate over all intermediate boundary energies.

\end{enumerate}

For an example of application of these rules, see Figure \ref{fig:resolve_rules}. We note that applying the rules to the diagram obtained by closing the boundary legs of $\ker[\corr{{\rm I}\hat{\calO}_{n}{\rm I}_{E_n}\dots\hat{\calO}_{1}{\rm I}_{E_1}}] $, we reproduce the black hole correlator $\bcorr{\tr({\rm P} \,\hat{\calO}_{n}{\rm I}_{E_n}\dots \hat{\calO}_{1}{\rm I}_{E_1})}$. This is because for an operator $\Psi[f]$ with $f=f(E)$, $\tr(\Rho\, \Psi[f])=\int dE \int d\mu\,  \rho_{\text{Pl}}\cdot 8 b\cdot f = \int dE \, \rho_{\Sch}\cdot f$. 



\section{Discussion}

We performed a group-theoretic analysis of correlators of boundary operators in the Schwarzian limit of JT gravity, possible due to our construction of Hilbert spaces for boundary and matter consisting of wavefunctions with support on $\tAdS_2$, organized into irreps of $\tSL(2,\RR)$. This revealed a basic unit of quantum dynamics in the JT black hole, a transition amplitude between bulk and boundary emission of matter. 

Our analysis focused on generic components of correlators that do not depend on details of the interaction patterns of bulk fields. However, it would be interesting to study simple but non-trivial examples of local interactions of bulk fields, and see how they manifest themselves in boundary correlators; it may be possible to use our transition amplitudes to map bulk dynamics onto the boundary.

We also note that our analysis rested heavily on the $\tSL(2,\RR)$ symmetry of $\tAdS_2$. Meanwhile, exponentially accurate contributions to correlators of the Schwarzian theory originate from spacetimes with non-trivial topology \cite{Saad19}, which no longer possess this symmetry. How the Hilbert space for JT gravity on $\tAdS_2$ can be extended to include fluctuations of boundaries of spacetime with non-trivial topology is an open question.

\section*{Acknowledgments}
We thank Alexei Kitaev, Joaquin Turiaci, Herman Verlinde, and Zhenbin Yang for related discussions. We gratefully acknowledge support by the Simons Foundation through the ``It from Qubit'' program. This research was also supported in part by the National Science Foundation under Grant No. NSF PHY-1748958. We are especially thankful to Donald Marolf and the high-energy and gravity group at UCSB for hospitality while this work was being completed.
\appendix

\section{Curves in two dimensions}\label{app:curves}

For any vector $V$ with constant normalization along a curve $X(s)$ with proper parametrization, 
\be
\dot{X}^{\nu}\nabla_{\nu}V^{\mu}=A N^{\mu}
\ee
where $N$ is the unit normal to $V$. We may fix the orientation of $N$ to be 
\be
\left\{\begin{matrix}
\figbox{1}{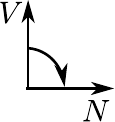}\\
\figbox{1}{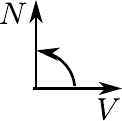}
\end{matrix} \,\right\}
\quad \text{for} \quad 
\left\{\begin{matrix}
\text{time-like}\\
\text{space-like}
\end{matrix}\right\} \quad V,
 \ee
 and define $\big[\dot{X}^{\nu}\nabla_{\nu} V^{\mu}\big]\equiv A$. Then the curvature of the curve is given by
 \be
 \kappa=\big[ \dot{X}^{\nu}\nabla_{\nu}\dot{X}^{\mu}\big], \qquad \dot{X}^{\nu}\nabla_{\nu}\dot{X}^{\mu}=\kappa n^{\mu}
 \ee
 where $n$ is the unit normal vector on a curve with orientation specified relative to $\dot{X}$ as above. Note the extrinsic curvature, defined for hypersurfaces in general, coincides with the above curvature. That is,
 \begin{align}
 \tilde{\kappa}&=h^{ab}e_{a}^{\al}e_{b}^{\beta}\nabla_{\al}n_{\bt}\nn
 &=h^{ss}\dot{X}^{\al}\dot{X}^{\bt}\nabla_{\al}n_{\beta}=-h^{ss}\dot{X}^{\alpha}\nabla_{\alpha}\dot{X}^{\beta}n_{\beta}=-h^{ss}\kappa n^{\bt}n_{\bt}=\kappa
  \end{align}
  where $a,b,\dots$ and $\al, \bt,\dots$ label intrinsic and external coordinates respectively, and $e_{a}^{\al}={\p x^{\al} \ov \p y^{a}}$ etc.
  
  Now, let $\{e_0, e_1\}$ be frame vectors on a $2$-dimensional Lorentzian manifold with $e_0, e_1$ respectively time-like and space-like and $e_1$ clock-wise oriented from $e_0$, $e^{a}_{\mu}e_{a\nu}=g_{\mu\nu}$, $e_{a\mu}e_{b}^{\mu}=n_{ab}$. For a time-like curve, we may define the angle $\al$ between $e_0$ and $\dot{X}$,
\be \label{app1}
\dot{X}^{\mu}=\cosh \al (e_0)^{\mu}+\sinh \al (e_1)^{\mu}.
\ee
See Figure \ref{fig:kappa}.
Then 
\be \label{app2}
\kappa=\kappa_e + \dot{\al}
\ee
where $\ka_{e}=\big[\dot{X}^{\nu}\nabla_{\nu}(e_0)^{\mu}\big]$, $\dot{X}^{\nu}\nabla_{\nu}(e_0)^{\mu}=\ka_e (e_1)^{\mu}$. That is, the rate of change of $\dot{X}$ along the curve is given by the sum of the rate of change of $e_0$ and the rate of change of the angle between $e_0$ and $\dot{X}$. (For a space-like curve, \eqref{app1}, \eqref{app2} hold with $e_0 \leftrightarrow e_1$.) Furthermore, $\ka_e=(e_1)_{\mu}\dot{X}^{\nu}\nab_{\nu}(e_0)^{\mu}=\dot{X}^{\mu}\om_{\mu}=\dot{X}^{\mu}\om_{\mu}$ where $\om_{\mu}=(e_1)_{\nu}\nab_{\mu}(e_0)^{\nu}$ is related to the spin connection associated with the frame $\{e_0, e_1\}$ as 
\be
\om_{\mu}^{ab}=e_{\nu}^{a}\nab_{\mu}e^{b\nu}=\om_{\mu}\Xi^{ab}
\ee
where $\Xi^{a}_{\,\,b}=\begin{pmatrix}
0 & 1\\
1 & 0
\end{pmatrix}$ is the boost matrix acting on vectors in the Lorentzian manifold. This results in the following formula for the curvature of a curve involving a frame field and its spin connection,
\be 
\kappa=\om_{\mu}\dot{X}^{\mu}+\dot{\al}.
\ee

\section{Curves in $\tAdS_2$}\label{app:AdS2_curves}

Here we apply the expression \eqref{kappa} to curves in $\tAdS_2$ to derive the Schwarzian limit of the JT action as well as equations of motion and solutions outside of the Schwarzian limit, which were presented in section $2$ of \cite{KiSuh18}.

Let us use global coordinates $(\phi, \tht)$ in which the metric is \eqref{AdSg} and non-zero Christoffel symbols are $\Ga^{\phi}_{\phi \tht}=\Ga^{\phi}_{\tht\phi}=\Ga^{\tht}_{\phi\phi}=\Ga^{\tht}_{\tht\tht}=\tan\tht$. A convenient frame field is 
\be\label{tildeg}
\left(e_{0}\right)^{\mu}=\de_{\phi}^{\mu}\cos\tht, \quad \left(e_1\right)^{\mu}=\de_{\tht}^{\mu}\cos\tht \hspace{80pt} \figbox{1}{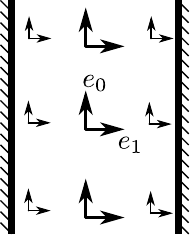}
\ee
for which $\om_{\phi}=\tan \tht$, $\om_{\tht}=0$. We consider time-like curves.

\subsection{Curvature}
We have 
\be
\dot{\phi}=\cos\tht \cosh \al, \quad \dot{\tht}=\cos \tht \sinh\al
\ee and $\dot{\al}=\vep_{\mu\nu}\dot{X}^{\mu}\ddot{X}^{\nu}$ where $\vep_{\phi \tht}=-\vep_{\tht\phi}=\sqrt{-g}$, so
\be
\ka=\om_{\mu}\dot{X}^{\mu}+\vep_{\mu\nu}\dot{X}^{\mu}\ddot{X}^{\nu}.
\ee
In the Schwarzian limit, $\de={\pi \ov 2}\mp \tht$, $|\al| \ll 1$, 
\begin{align}
\ka&=\sin\tht \cosh \al + \dot{\al}\nn
&\approx \pm \left(1-{\de^2 \ov 2}\right)\left(1+{\al^2 \ov 2}\right)+\dot{\al} \quad \left(\sin\tht=\pm \sin\left({\pi \ov 2}-\de\right)\approx \pm \left(1-{\de^2 \ov 2} \right), \,   \cosh\al \approx 1+{\al^2 \ov 2}\right)\nn
&=\pm \left(1-\left({1 \ov 2}\dot{\phi}^2 + {\phi^{(3)} \ov \dot{\phi}} -{3 \ov 2}\left({\ddot{\phi} \ov \dot{\phi}}\right)^2\right)\right)
 \quad \left(\de\approx |\dot{\phi}|,\,  \al\approx {\mp \ddot{\phi} \ov \dot{\phi}}\right).
 \end{align}
Thus if we define the extrinsic curvature $K$ in terms of the outward normal vector on each of $\calM_{\rm R}$ and $\calM_{\rm L}$ [fig]
\be \label{KSch}
K=\pm \ka = 1-\Sch\left(\tan{\phi(s) \ov 2}, s\right), \qquad \Sch\left(\tan{\phi(s) \ov 2}, s\right)={1 \ov 2}\dot{\phi}^2 + {\phi^{(3)} \ov \dot{\phi}} -{3 \ov 2}\left({\ddot{\phi} \ov \dot{\phi}}\right)^2
\ee
where the $+$ sign applies to a particle with $\dot{\phi}>0$ on $\calM_{\rm R}$ or $\dot{\phi}<0$ on $\calM_{\rm L}$, and the $-$ sign to the opposite situation.

\subsection{Equations of motion}
Let us consider the boundary action obtained from JT gravity with a Lagrange multiplier fixing total proper time,
\begin{align}
I_{R=-2}[X]&={\Phi_* \ov 2 \pi}\int ds \sqrt{-h}(K-1) + E_g \int ds \sqrt{-h}\nn
&\pm \ga \int \left(dX^{\mu}\om_{\mu}+d \al\right)- M \int dl, \qquad \ga={\phi_* \ov 2 \pi}, \quad M=\ga-E_g
\end{align}
where the signs were explained below \eqref{kappa} and again below \eqref{KSch} as applied to the Schwarzian limit.
The Lagrangian w.r.t. an arbitrary time parameter is 
\begin{align}
\calL_{R=-2}&=\pm \ga \left( \om_{\mu}\dot{X}^{\mu}+\dot{\al}\right)-M \dot{\ell}\nn
&=\pm \ga\left(\om_{\mu}\dot{X}^{\mu}+{\vep_{\rho\sig}\dot{X}^{\rho}\ddot{X}^{\sig} \ov \left(-g_{\mu\nu}\dot{X}^{\mu}\dot{X}^{\nu}\right)}\right)-M \sqrt{-g_{\mu\nu}\dot{X}^{\mu}\dot{X}^{\nu}}.
\end{align}
After varying $\calL_{R=-2}[X]$, it is convenient to fix the time parameter to be the proper time of the unperturbed trajectory. For the variational problem to be well-defined, we need at endpoints of the particle trajectory
\be \label{Lag}
\de \al =\vep_{\mu\nu}\dot{X}^{\mu}\de \dot{X}^{\nu}=0 \qquad \text{and} \qquad \de X^{\mu} \propto -\ga \tan \tht \dot{\tht}\dot{X}^{\mu}+\left(\ga \tan \tht \dot{\phi} - M\right)n^{\mu}.
\ee
Then in the variation of \eqref{Lag} we can subtract the variation of $\dot{\al}$ which integrates to zero to find 
\be
\de \calL_{R=-2}=\left( \pm \ga \left( \p_{\mu}\om_{\nu}-\p_{\nu}\om_{\mu}\right)\dot{X}^{\nu}-M \ka n_{\mu}\right)\de X^{\mu}.
\ee
The resulting equation of motion is $\left(\de \calL_{R=-2}/ \de X^{\mu}\right)n^{\mu}=\pm \ga-M \ka=0$, or
\be \label{eom}
\ga=M K.
\ee
For completeness we note an identity that is useful in calculations involving the $R=-2$ action,
\be
{d \ov ds}\left(g_{\mu\nu}\dot{X}^{\mu}\dot{X}^{\nu}\right) \quad \Rightarrow \quad \vep_{\mu\nu}\dot{X}^{\mu}\ddot{X}^{\nu}=-\tan \tht \dot{\tht}.
\ee

\subsection{Solutions to equations of motion}
It is convenient to use the ambient space of $\AdS_2$ with coordinates $\{\calX^{0} , \calX^1, \calX^{2}\}$ and metric $ds^2=-\left(d\calX^1\right)^2+\left(d\calX^1\right)^2-\left(d\calX^1\right)^2$ and in which the unit normal to the $\AdS_2$ surface
\be
\calX^{0}={\cos \phi \ov \cos\tht}, \quad \calX^{1}=\tan \tht, \quad \calX^{2}={\sin \phi \ov \cos \tht}
\ee
 is given by $\calX^{A}$, $\calX^{A}\calX_{A}=-1$. 

Given \eqref{eom}, we may solve for
\be \label{eompair}
\dot{X}^{\nu}\nabla_{\nu}\dot{X}^{\mu}=K N^{\mu}, \qquad \dot{X}^{\nu}\nabla_{\nu}N^{\mu}=K \dot{X}^{\mu}.
\ee
Lifting the vector $\dot{X}^{\mu}$ to the ambient space as $\dot{\calX}^{A}=e_{\mu}^{A}\dot{X}^{\mu}$, $\dot{X}_{\mu}=e_{\mu}^{A}\dot{X}_{A}$ and similarly $N^{\mu}$ to $\calN^{A}$, we find using \eqref{eompair} and ${d \ov ds}\left(\dot{\calX}^{A}\calX_{A}\right)={d \ov ds}\left(\dot{\calN}^{A}\calX_{A}\right)=0$ that
\be
\ddot{\calX}^{A}-K \calN^{A}=\vep \calX^{A}, \qquad \dot{\calN}^{A}-K \dot{\calX}^{A}=0
\ee
where $\vep =\dot{\calX}^{A}\dot{\calX}_{A}= \pm 1 $ with the upper (lower) sign for space-like (time-like) trajectories. In particular, $\calQ^{A}=\calN^{A}-K \cal{X}^{A}$ is a constant vector such that
\be
\calQ^{A}\calX_{A}=K, \qquad \calQ^{A}\calQ_{A}=-\vep-K^2,
\ee
and classical trajectories that are not null are given by the intersection of a hyperplane normal to $\calQ^{A}$ with $\AdS_2$.

\section{Intertwiners, $6j$-symbols, and matrix elements of interactions}\label{app:calc}
As explained in Section \ref{sec:scheme}, our general strategy for deriving \eqref{diag1}-\eqref{diag5} with \eqref{rho} and \eqref{rhom}-\eqref{Wilp} is to utilize the embedding \eqref{embed}
\be \label{embedA}
\Xi^{\mu \pm}_{\lam}: \calU^{\mu}_{\lambda} \to  \calF^{\mu}_{\lam},\qquad  \Xi^{\mu \pm}_{\lam}\ket{m}=c^{\pm}_{\lam, m}f_{\lam, m}
\ee
to i) evaluate a diagram with respect to spaces $\calF^{\mu}_{\lam}$, then if necessary ii) obtain the commuting map involving irreps $\calU_{\lam}^{\mu}$, and finally iii) dress the resulting expression with overall coefficients associated with each three-way vertex, coming from relative factors of matrix elements of interactions over Clebsch-Gordon coefficients. In calculating matrix elements of interactions, we will use boundary wavefunctions in $\calH^{\pm}$ rather than $\widetilde{\calH}^{\pm}$, see Footnote \ref{ospin}.

We present relevant calculations below. As a preliminary, let us discuss some aspects of \eqref{embedA}. The coefficients $c^{\pm}_{\lam, m}$ are given by
\be \label{coeff}
c^{\pm}_{\lam, m}=\sqrt{\Ga(\lam \pm m) \ov \Ga(1-
\lam \pm m)}(\pm 1)^{m-\mu}
\ee
When $\calU^{\mu}_{\lam}$ is a principal irrep $\calC^{\mu}_{\lam}$, both embeddings exist with coefficients differing by an overall factor. Furthermore, composing $\Xi^{\mu \pm}_{\lam}$ with its conjugate $\Xi_{\lam}^{\mu\pm \dagger}: \calF_{1-\lam}^{\mu} \to \calC^{\mu}_{\lam}$ with respect to the inner product 
\be
\corr{g| f}=\int_{0}^{2\pi}{d\vp \ov 2 \pi}\, g^*(\vp)f(\vp) \quad \text{where}\quad g\in \calF^{\mu^*}_{1-\lam^*}, \, f \in \calF_{\lam}^{\mu}
\ee
we obtain an isomorphism $\Sig_{\lam}^{\mu \pm}=\Xi_{\lam}^{\mu\pm}\Xi_{\lam}^{\mu\pm \dagger}: \calF_{1-\lam}^{\mu} \to \calF_{\lam}^{\mu}$. In our calculations we use although in a minor way its position space kernel
\be
\Sig_{\lam}^{\mu\pm}(\vp_1, \vp_2)={2\pi i \ov \Ga(1-2\lam)}{\pm e^{i \pi \lam} \ov e^{2\pi i (\mu \pm \lam)}-1}\begin{dcases}(\vp_{12})^{-2\lam} & (\vp_2+2\pi > \vp_1 > \vp_2)\\
e^{- 2\pi i \mu}(\vp_{21})^{-2\lam} & (\vp_2 > \vp_1 > \vp_2-2\pi)
\end{dcases}.
\ee
For a discrete irrep $\calD^{\pm}_{\lam}$, only the embedding with same sign exists, and we denote the image of the embedding by $\calF_{\De}^{\pm}\subset \calF_{\De}^{\pm \De}$.

\subsection{Scalar wavefunctions on $\tAdS_2$} \label{app:scwf}

Here we give wavefunctions which will be used in propagators for a scalar field of dimension $\De>1/2$. They are obtained by taking $\nu \to 0$ in the spinor solutions described in Appendix A.3 of \cite{KiSuh18}. With the normalization convention given under \eqref{twop}, wavefunctions corresponding to states of an irrep $\calD^{\mp}_{\lam}$ are given by
\be \label{psi0}
\psi^{0}_{\lam, \mp(\lam+k)}(\phi, \tht)={\Ga(\lam) \sqrt{\lam-1/2} \ov 2 \pi}\sqrt{{\Ga(2\lam+k)} \ov k!}i^{\pm k}(-u)^{-\lam-k \ov 2}(1-u)^{\lam}\hgfs(-k, \lam,2\lam; 1-u)e^{\mp i(\lam+k)\phi}
\ee
where $\hgfs$ denotes the regularized hypergeometric function and $u=e^{i (\pi-2 \tht)}$. Note $\left(\psi^0_{\lam, \mp (\lam+k)}\right)^*=(-1)^k \psi^0_{\lam, \pm (\lam+k)}$. When integrating against boundary wavefunctions in $\calH^{\pm}$, only the behavior of \eqref{psi0} near the asymptotic boundaries of $\tAdS_2$ is relevant:
\be \label{psi0asym}
\psi^{0}_{\lam, \mp(\lam+k)} \underset{\tht \to \pm {\pi \ov 2}}{\approx}{\sqrt{\lam-1/2} \ov 2 \pi}{\Ga(\lam) \ov \Ga(2\lam)}(-1)^k \sqrt{{\Ga(2\lam+k)  \ov k!}}\left({\upsilon \ov \ga}\right)^{\lam}e^{\mp i (\lam+k)\phi}
\ee
where we have used the asymptotic coordinate $\upsilon$ defined \eqref{asymc}. The scaling with $\ga \sim \ep^{-1}$ in the radial coordinate implies we should renormalize boundary operators creating these fields as \eqref{renorm}.

\subsection{Intertwiners and matrix elements of interactions} \label{app:inter}

\subsubsection{Involving one discrete irrep}

Let us first consider an intertwiner involving one discrete irrep, $V: \calF^{\mu_1}_{\lam_1} \to \calF_{\lam_2}^{\mu_2} \otimes \calF_{\De}^{\mp}$. For convenience, we choose to define the commuting intertwiner $\Ga: \calC^{\mu_1}_{\lam_1} \to \calC^{\mu_2}_{\lam_2} \otimes \calD^{\mp}_{\De}$ using the following combination of embeddings:
\be \label{commapp}
\figbox{1.1}{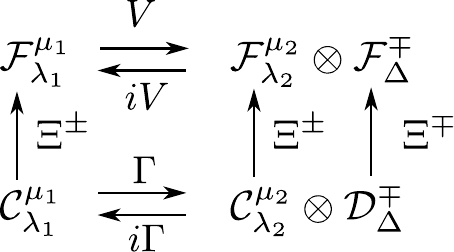}\quad  .
\ee
The position-space kernel of $V_{\lam_1, \mu_1}^{\lam_2, \mu_2;\De, \mp}$ can easily be determined up to normalization from symmetry considerations; it belongs to the one-dimensional space of invariants in $\calF^{\mu_2}_{\lam_2}\otimes \calF^{\mp}_{\De}\otimes \calF^{-\mu_1}_{1-\lam_1}$. (See Section 6 of \cite{SL2R} for the argument that this space is one-dimensional.) We choose a normalization such that $V_{\lam_1, \mu_1}^{\lam_2, \mu_2;\De, \mp}\Sig_{\lam_1}^{\mu_1\pm} \propto V_{1-\lam_1, \mu_1}^{\lam_2, \mu_2;\De, \mp}$ with an overall factor that only depends on $(\lam_1, \mu_1)$, and similarly for acting with $\Sig_{\lam_2}^{\mu_2 \pm}$ after $V_{\lam_1, \mu_1}^{\lam_2, \mu_2;\De, \mp}$. This will ensure that coefficients of the associator
\be \label{assrel}
\left[\figbox{0.66}{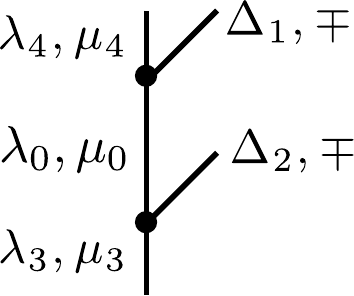}\right]_{\calF}=\sum_n F^{\lam_4, \De_1, \De_2, \mp}_{\lam_3}(\lam_0, \mu_0; n) \left[\figbox{0.66}{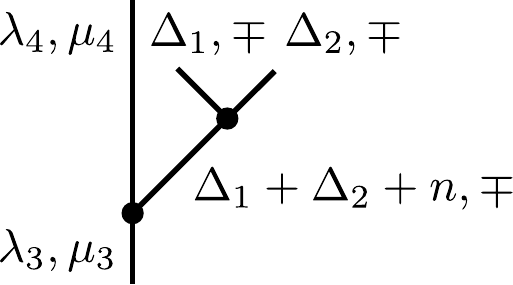}\right]_{\calF}
\ee
 (where the subscript $\calF$ indicates that diagrams are evaluated with respect to Hilbert spaces $\calF^{\mu}_{\lam}$ and in particular that the intertwiner $V$ is used on vertices) are invariant under taking $\lam  \to 1-\lam$. 
 
 We fix the position-space kernel of $V$ to be (for $\calF_{\De}^{\mp}$ we use the position-space coordinate $z=e^{i\vp}$)
 \begin{align} \label{Vfunc}
&V^{\lam_2, \mu_2, \De, \mp}_{\lam_1, \mu_1}(\vp_2, z; \vp_1)\nn
&=\Ga(\De+\lam_1-\lam_2)^{-1}z^{\mp \De} \left(e^{\mp i \vp_2/2}-z^{\mp 1 }e^{\pm i \vp_2/2} \right)^{1-\lam_1-\lam_2-\De} \left( e^{\mp i \vp_1/2}-z^{\mp 1}e^{\pm i \vp_1/2}\right)^{\lam_1+\lam_2-1-\De}\nn
&\begin{dcases}
\vp_{21}^{\De+\lam_1-\lam_2-1}& (\vp_1+2\pi > \vp_2 > \vp_1)\\
e^{\pm i \pi(\lam_1+\lam_2-1-\De)}e^{-2\pi i \mu_1}\vp_{12}^{\De+\lam_1-\lam_2-1} & (\vp_1> \vp_2>\vp_1-2\pi)
\end{dcases}.
\end{align}
Fourier-transforming and using \eqref{commapp} and \eqref{coeff}, we obtain Clebsch-Gordon coefficients of the intertwiner $\Ga$ commuting with $V$,
\begin{align} \label{inter1}
\left[\figbox{0.7}{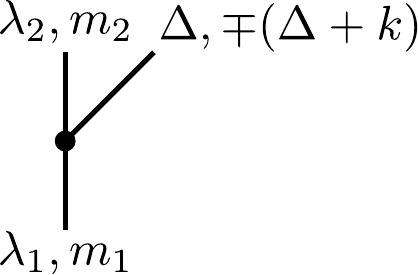}\right]_{\calU}&=c_{\lam_1, m_1}^{\pm}\left(c^{\pm}_{\lam_2, m_2}\right)^{-1}\left(c^{\mp}_{\De, \mp(\De+k)}\right)^{-1}\left[\figbox{0.7}{intertwiner_F}\right]_{\calF}\nn
&=i^{\pm(\lam_1+\lam_2-1-\De)}\sqrt{{ \Ga(\lam_1 \pm m_1)\Ga(1-\lam_1 \pm m_1)\ov \Ga(\lam_2\pm m_2)\Ga(1-\lam_2 \pm m_2) }}\underbrace{\left( {\sin \pi (\lam_1\mp m_1) \ov \pi}e^{-i \pi m_1}\right)}_{\mp e^{i \pi \lam_1}\left(1-e^{\pm 2 \pi i (\lam_1\mp \mu_1)}\right)}\nn
&\times {{(-1)^k S_{k}\left(s_1, \De+i s_2, \De-i s_2, {1 \ov 2}\pm m_1\right)} \ov \sqrt{\Ga(2\De+k) k!}}, \qquad k=\mp(m_1-m_2)-\De.
\end{align}
where
\be
S_{n}(x; a, b, c)=(a+b)_n(a+c)_n\, {}_{3}F_{2}
\left( \begin{matrix}
-n, & a+i x, & a-i x\\
a+b, &a+c
\end{matrix}; 1\right)
\ee
is the continuous dual Hahn polynomial of $n'$th degree in $x^2$, symmetric in $a, b, c$. We note $S_{k}\left(s_1; \De+i s_2, \De-i s_2, {1 \ov 2} \pm m_1\right)=(-1)^k S_{k}\left(s_2; \De+i s_1, \De-i s_1, {1 \ov 2}\mp m_2\right)$. This is the same function as appears in Clebsch-Gordon coefficients obtained in Section 3.2 of \cite{Gro05}. Using large-$k$ asymptotics of the function $S_{k}$, we find the norm of $\Ga$ to be
\be \label{bnorm}
\left[\figbox{0.7}{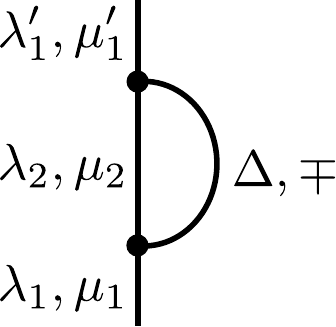}\right]_{\calF, \calU}=\de(E_1-E_1')N_{\lam_1, \mu_1}^{\lam_2, \De, \mp}, \quad N_{\lam_1, \mu_1}^{\lam_2, \De, \mp}={e^{\pm i \pi(\lam_1+\lam_2-1)} \ov 8 \pi^2} \rho_{\text{Pl}}(E_1, \mu_1)^{-1} \Ga(\De \pm i s_1 \pm is_2)^{-1}.
\ee
Note the norm of $\Ga$ and $V$ are in fact the same, as the coefficients involved in the commutation relation between intertwiners \eqref{commapp} cancel between the two external legs in the diagram of \eqref{bnorm}.

Finally, let us obtain matrix elements of interactions. Using the integral formula for boundary wavefunctions \eqref{psi}
\be
\psi^{\pm}_{\lam, m}(\phi, \upsilon)=e^{i m \phi} \sqrt{2\pi \rho_{\text{Pl}}(E, \mu) {\Ga(1-\lam \mp m) \ov \Ga(\lam \mp m)}}e^{\pm i \pi m}\upsilon^{\lam}\int_{0}^{2\pi}{dt \ov 2 \pi}{e^{\mp i m t+i {\ups \ov 2}\cot{t \ov 2}} \ov \left(2\sin{t \ov 2}\right)^{2\lam}}
\ee
which amounts to a decomposition using bulk-boundary propagators, and the asymptotic behavior of matter functions \eqref{psi0asym}, we obtain
\begin{align} \label{bmel}
\figbox{0.75}{intertwiner_F}&=\int_{\calM}\left(\psi^{\pm}_{\lam_2, m_2}\right)^*\cdot\ga^{\De}\left(\psi^{0}_{\De, \mp(\De+k)}\right)^*\cdot \psi^{\pm}_{\lam_1,m_1}\nn
&=\de(\pm(m_1-m_2)+\De+k)\,C^{\lam_2, \De \mp}_{\lam_1, \mu_1}\left[\figbox{0.7}{intertwiner_F}\right]_{\calU},
\end{align}
\be \label{bratio}
C^{\lam_2, \De \mp}_{\lam_1, \mu_1}={\Ga(\De) \sqrt{\De-{1 \ov 2}} \ov 2 \pi}{e^{i \pi m_1} \ov \sin\pi(\lam_1\mp m_1)}{i^{\mp (\lam_1+\lam_2-1-\De)} \ov \sqrt{\Ga(2\De)}}\Ga(\De \pm is_1 \pm is_2)\sqrt{\rho_{\Sch}(E_1)\rho_{\Sch}(E_2)}.
\ee
Using \eqref{bnorm} and \eqref{bratio} we derive \eqref{diag1},
\be \label{reabsorbamp}
\figbox{0.75}{reabsorb_app}=\de(E_1-E'_1)N_{\lam_1, \mu_1}^{\lam_2, \De, \mp}\left |C^{\lam_2, \De \mp}_{\lam_1, \mu_1}\right |^2=\de(E_1-E'_1)\rho_{\Sch}(E_2)\rho_m(\De)A_{\bdy}(E_1, E_2, \De).
\ee
Note in \eqref{reabsorbamp} the dependence on $\mu_1$ in \eqref{bratio} cancels against that in \eqref{bnorm}---this would not be the case if we had used boundary wavefunctions $\psi^{\mp}$ rather than $\psi^{\pm}$ in \eqref{bmel}. Similarly, the absence of dependence on $\mu$ parameters in \eqref{diag2} as well as in the coefficients in \eqref{diag3}-\eqref{diag5} depend on \eqref{tHS}.
\subsubsection{Involving three discrete irreps}
Now we consider an intertwiner between three irreps, $V: \calF^{\mp}_{\lam} \to \calF_{\lam_1}^{\mp} \otimes \calF_{\lam_2}^{\mp}$. It is convenient to consider the larger commutative diagram
\be
\figbox{1.1}{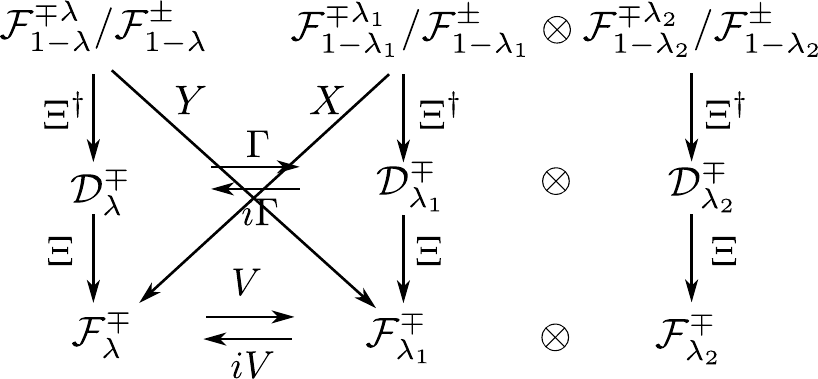}
\ee
and define $V$, $iV$ by commutation with $Y$, $X$. (See \cite{SL2R} for a description of the quotient spaces appearing above.) The space of intertwiners is again one-dimensional, and we fix the position space kernel of $Y$ to be
\be
Y^{\mp}_{\lam_1, \lam_2; \lam}(z_1, z_2; w)=z_1^{\mp \lam_1}z_2^{\mp \lam_2}w^{\pm \lam}\left( z_2^{\mp 1}-z_1^{\mp 1}\right)^{n}\left(1-(z/w_1)^{\mp 1}\right)^{-2\lam_1-n}\left(1-(z/w_2)^{\mp 1}\right)^{-2\lam_2 -n}
\ee
where
\be
\lam=\lam_1+\lam_2+n.
\ee
Then the commuting $X$ is given by
\be
X^{\mp}_{\lam; \lam_1, \lam_2}(z; w_1, w_2)=z^{\mp \lam}w_1^{\pm \lam_1}w_2^{\pm \lam_2}\left( w_2^{\pm 1}-w_1^{\pm 1}\right)^{n}\left(1-(w/z_1)^{\pm 1}\right)^{-2\lam_1-n}\left(1-(w/z_2)^{\pm 1}\right)^{-2\lam_2 -n}
\ee
and we can derive a functional form for $iV$
\be \label{iV}
\mathring{\left(iV^{\mp}_{\lam; \lam_1, \lam_2}f\right)}(z^{\mp})=(\mp 1)^n n! \sum_{n_1+n_2=n}\left.{1 \ov \Ga(2\lam_1+n_1)\Ga(2\lam_2+n_2)}{(-\p_1)^n \ov n_1!}{(\p_2)^n \ov n_2!}\mathring{f}\left(z_1^{\mp 1}, z_2^{\mp 1}\right)\right|_{z_1=z_2=z}
\ee
with $\mathring{f}(z)=e^{\pm i \lam \vp}f(z)$ for $f \in \calF^{\mp}_{\lam}$ which will be useful for deriving a $6j$-symbol in the next section, as well as Clebsch-Gordon coefficients of $i\Ga$
\begin{align}
&\left[\figbox{0.7}{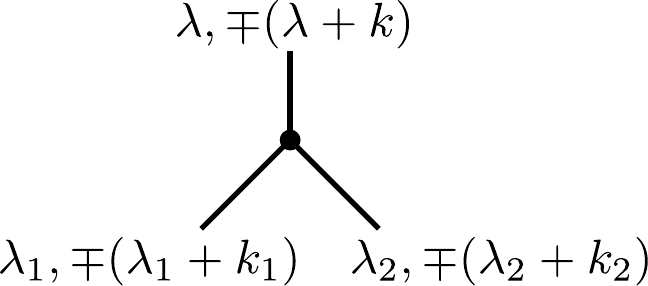}\right]_{\calU}={ c_{\lam_1, \mp(\lam_1+k)}^{\mp} c_{\lam_2, \mp(\lam_2+k)}^{\mp} \ov c_{\lam, \mp(\lam+k)}^{\mp}}\left[\figbox{0.7}{disc_opp_intertwiner}\right]_{\calF}\nn
&=n!(\mp 1)^n\sqrt{{k! \ov \Ga(2\lam+k)}}\sqrt{{k_1! \ov \Ga(2\lam_1+k_1)}}\sqrt{{k_2! \ov \Ga(2\lam_2+k_2)}}\sum_{\substack{n_1+n_2=n\\
 n_1 \leq k_1, n\\
  n_2 \leq k_2, n}}{(2\lam_1 +n)_{k_1-n_1} \ov n_1! (k_1-n_1)!}{(2\lam_2 +n)_{k_2-n_2} \ov n_2! (k_2-n_2)!}(-1)^{n_1}.
\end{align}
The norm of $\Ga$ was given in \cite{SL2R},
\be \label{mnorm}
\left[\figbox{0.7}{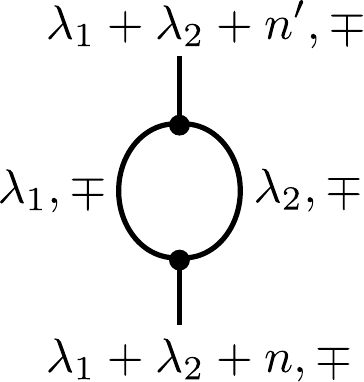}\right]_{\calF, \calU}=\de_{n,n'}d_n, \quad d_n={n! \ov (2\lam-1)\Ga(2\lam_1+2\lam_2+n-1)\Ga(2\lam_1+n)\Ga(2\lam_2+n)}
\ee

Next we obtain matrix elements using wavefunctions \eqref{psi0} which are finite sums:
\begin{align} \label{mmel}
\figbox{0.75}{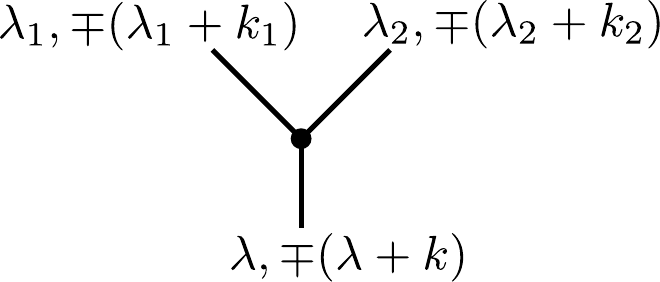}&=\int_{\tAdS_2}\left(\psi^{0}_{\lam_1, \mp(\lam_1+k_1)}\right)^*\cdot\left(\psi^{0}_{\lam_2, \mp(\lam_2+k_2)}\right)^*\cdot\psi^{0}_{\lam, \mp(\lam+k)}\nn
&=\de(\lam_1+k_1+\lam_2+k_2-(\lam+k))\,C^{\lam_1, \lam_2}_{n}\left[\figbox{0.7}{disc_intertwiner}\right]_{\calU},
\end{align}
\be \label{mratio}
C_{n}^{\lam_1, \lam_2}=\sqrt{\left(\lam-{1 \ov 2}\right)\left(\lam_1-{1 \ov 2}\right)\left(\lam_2-{1 \ov 2}\right)}{2^{2\lam} \ov 4 \pi n!}{\Ga(\lam_1+\lam_2+{n-1 \ov 2}) \ov \Ga({1-n \ov 2})}\Ga\left(\lam_1+{n \ov 2}\right)\Ga\left(\lam_2+{n \ov 2}\right).
\ee
Note the matrix elements vanish for odd $n$. Using \eqref{mnorm} and \eqref{mratio} we derive \eqref{diag2},
\be \label{remergeamp}
\figbox{0.75}{merge_app}=\de_{n,n'}d_n \left |C^{\lam_1, \lam_2}_{n}\right |^2=\de_{n,n'}\rho_m(\lam_1)\rho_m(\lam_2)A_{\blk}(\lam_1, \lam_2, n).
\ee

\subsection{$6j$-symbols} \label{app:6j}

Using results obtained in Section \ref{app:inter}
we can calculate the $6j$-symbols relevant to our problem, which occur in the relations \eqref{uncross}, \eqref{ass}, and \eqref{diss}. 

Let us first solve the uncrossing relation \eqref{uncross}. The first step is to solve for the relation in which diagrams have been replaced with ones with respect to spaces $\calU_{\lam}^{\mu}$,
\be \label{uncrossF}
\left[\figbox{0.66}{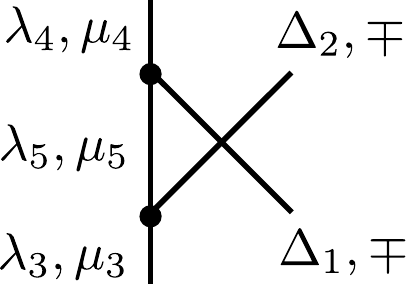}\right]_{\calU}\quad = \int dE_6 \, F^{\lambda_4, \mu_4, \Delta_2, \mp}_{\lambda_3, \mu_3, \De_1, \mp}(\lambda_5; \lambda_6)\quad \left[\figbox{0.66}{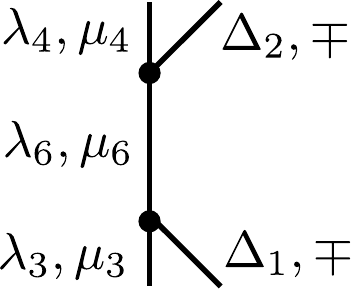} \right]_{\calU}.
\ee
After attaching the intertwiner $\Ga_{\lam_0, \mu_0}^{\lam_3, \mu_3,\De_1, \mp}$ to both sides and using the norm of $\Ga$ \eqref{bnorm} on the RHS, we obtain 
\be
\left[\figbox{0.66}{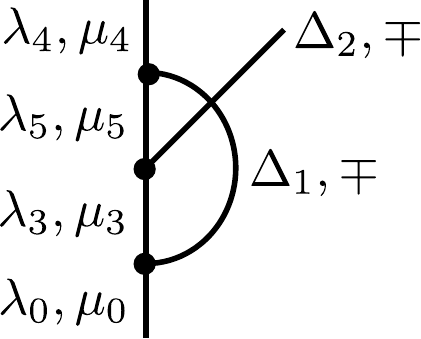}\right]_{\calU}\quad = F^{\lambda_4, \mu_4, \Delta_2, \mp}_{\lambda_3, \mu_3, \De_1, \mp}(\lambda_5; \lambda_0) \,N_{\lam_0, \mu_0}^{\lambda_3, \De_1, \mp}\quad \left[\figbox{0.66}{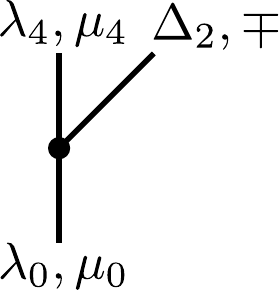} \right]_{\calU}.
\ee
Showing only momenta, 
\be
\left[\figbox{0.66}{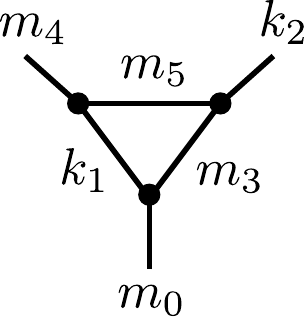}\right]_{\calU}= F^{\lambda_4, \mu_4, \Delta_2, \mp}_{\lambda_3, \mu_3, \De_1, \mp}(\lambda_5; \lambda_0) \,N_{\lam_0, \mu_0}^{\lambda_3, \De_1, \mp}\quad \left[\figbox{0.66}{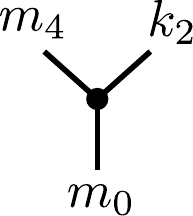} \right]_{\calU}.
\ee
The sum on the left-hand-side
\be
\text{LHS}=\sum_{k_1=0}^{\infty} i\Ga^{\lam_4, \mu_4}_{\lam_5,\mu_5, \De_1, \mp}(m_4; m_5, k_1)\, \Ga_{\lam_3, \mu_3}^{\lam_5, \mu_5, \De_2, \mp}(m_5, k_2;  m_3)\, \Ga^{\lam_3, \mu_3, \De_1, \mp}_{\lam_0, \mu_0}(m_3, k_1; m_0)
\ee
where matrix elements of $\Ga$ are given by \eqref{inter1} and those of the inverse $i\Ga$ are obtained by complex 
conjugation (as $\Ga$ is unitary), can be performed with the formula given in Theorem 7.3 of \cite{Gro05}. We find
\begin{align} \label{bareuncross}
F^{\lambda_4, \mu_4, \Delta_2, \mp}_{\lambda_3, \mu_3, \De_1, \mp}(\lambda_5; \lambda_6)&=e^{\mp i \pi(\lam_6-\lam_5)}e^{i \pi m_4}{\sin \pi(\lam_4 \pm \mu_4) \ov \pi}e^{-i \pi m_3}{\sin\pi(\lam_3 \mp m_3) \ov \pi}8 \pi^2 \rho_{\text{Pl}}(E_6, \mu_6) \nn
&\times\Ga(\De_1 \pm is_3 \pm is_6)\Ga(\De_2 \pm is_4\pm is_6)\nn
&\times\phi_{s_3}\left( s_4; \lambda_6+\Delta_2-{1 \over 2},  \Delta_1+\lambda_5-{1 \over 2},  \Delta_1-\lambda_5+{1 \over 2},  \lambda_6-\Delta_2+{1 \over 2}\right)
\end{align}
where the Wilson function $\phi_{\lam}(x; a, b, c, d)$ was described in \eqref{Wilf}. Note that as was denoted in \eqref{inter1}, the expressions $\sin \pi (\lam\mp m)e^{- i \pi m}$ and $\sin \pi (\lam\pm m)e^{+i \pi m}$ are in fact only dependent on the $\mu$ parameter of the representation. To finally obtain coefficients in \eqref{uncross}, we use the relative coefficient between matrix elements of interactions in our Hilbert space and those of $\Gamma$ obtained in \eqref{bmel}, \eqref{bratio},
\be \label{ratiouse}
F^{E_4, \De_2}_{E_3, \De_1}(E_5;E_6)=F^{\lambda_4, \mu_4, \Delta_2, \mp}_{\lambda_3, \mu_3, \De_1, \mp}(\lambda_5; \lambda_6) {C_{\lam_3, \mu_3}^{\lam_5, \De_2, \mp}C_{\lam_4, \mu_4}^{\lam_5, \De_1, \mp *} \ov C_{\lam_6, \mu_6}^{\lam_4, \De_2, \mp} C_{\lam_6, \mu_6}^{\lam_3, \De_1, \mp *}}.
\ee
This results in \eqref{diag3} in which $\mu$-dependences in the bare $6j$-symbol in \eqref{bareuncross} have been eliminated and gamma functions factor out as interaction amplitudes, leaving the precise gravitational scattering amplitude \eqref{Samp}.

Next, we turn to solving the associator and dissociator \eqref{ass}, \eqref{diss}. Let us attach the intertwiner $iV^{\mp}_{\De_1+\De_2+n; \De_1, \De_2}$ (whose functional form we determined in \eqref{iV}) to both sides of \eqref{assrel}.\footnote{Note the coefficients in this relation are identical whether we interpret diagrams with respect to $\calF^{\mu}_{\lam}$ or $\calU^{\mu}_{\lam}$. This was also the case for \eqref{uncrossF}. Here we choose to calculate the coefficients using diagrams with respect to $\calF^{\mu}_{\lam}$, \ie using intertwiners $V$ and $iV$.}After using \eqref{mnorm} on the RHS, 
\be
\left[\figbox{0.66}{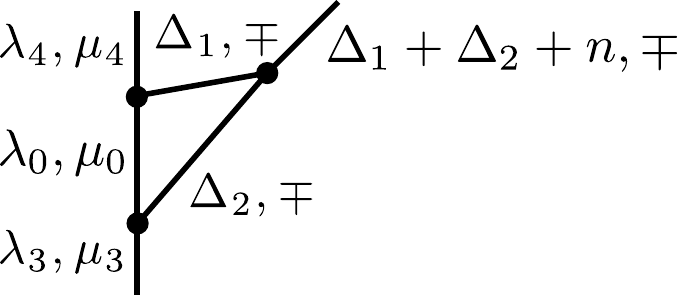}\right]_{\calF}\quad = F^{\lam_4, \De_1, \De_2, \mp}_{\lam_3}(\lam_0, \mu_0; n) \,d_n\quad \left[\figbox{0.66}{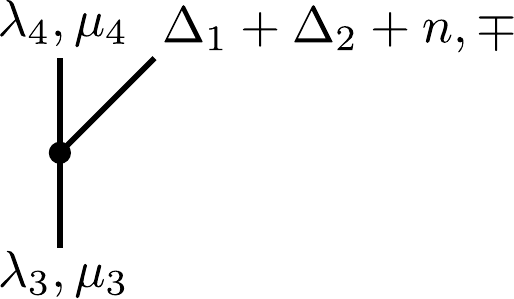} \right]_{\calF}.
\ee
By acting with $iV^{\mp}_{\De_1+\De_2+n; \De_1, \De_2}$ on
\begin{align}
&\left[\figbox{0.66}{unassoc_app}\right]_{\calF}={1-e^{\pm 2 \pi i (\lam_0 \mp \mu_0)} \ov 2 \pi}z_1^{\mp \De_1}z_2^{\mp \De_2}e^{\mp i \left({\lam_3 + \lam_4-1-\De_1 -\De_2 \ov 2} \right)\vp_3}e^{\mp i \left({-\lam_3 - \lam_4+1-\De_1 -\De_2 \ov 2} \right)\vp_4}\vp_{43}^{\lam_3-\lam_4-1+\De_1+\De_2}\nn
& \times\left(1-\left({z_1 \ov z_3}\right)^{\mp 1}\right)^{\lam_4-\De_1}\left(1-\left({z_1 \ov z_4}\right)^{\mp 1}\right)^{-\lam_4-\De_1}\left(1-\left({z_2 \ov z_3}\right)^{\mp 1}\right)^{\lam_3-1-\De_2}\left(1-\left({z_2 \ov z_4}\right)^{\mp 1}\right)^{-\lam_3+1-\De_2}\nn
&\times (1-\chi)^{\lam_0}\hgfs(\lam_0+\lam_3-1+\De_2, \lam_0-\lam_4 + \De_1, \lam_3 -\lam_4+\De_1+\De_2; \chi) \hspace{20pt}   (\text{for } \vp_3+2\pi > \vp_4 > \vp_3 )
\end{align}
where
\be
\chi={(z_1-z_2)(z_3-z_4) \ov (z_1-z_3)(z_2-z_4)}
\ee
which can be obtained from \eqref{Vfunc}, and comparing with the intertwiner $V_{\lam_3, \mu_3}^{\lam_4, \mu_4, \De_1+\De_2+n, \mp}$ again given by \eqref{Vfunc}, for example at $\vp_3=0, \vp_4=\pi$, we find
\begin{align}
F_{\lam_3}^{\lam_4, \De_1, \De_2, \mp}(\lam_0, \mu_0; n)&={1-e^{\pm 2 \pi i (\lam_0 \mp \mu_0)} \ov 2 \pi}(2 \De_1+2 \De_2 + 2 n -1)\Ga(2 \De_1 + 2 \De_2 + n-1) \nn
&\times{(\mp 1)^n \ov n!}W_{n}(s_0; \Delta_1+i s_3, \Delta_1-i s_3, \Delta_2+i s_4, \Delta_2-i s_4)
\end{align}
where $W_n(x; a, b,c, d)$ is the Wilson polynomial defined in \eqref{Wilp}. 

We now consider the inverse of \eqref{assrel}, 
\be \label{dissrel}
\left[\figbox{0.66}{assoc_app}\right]_{\calF}=\int dE_0 \left(F^{\lam_4, \De_1, \De_2, \mp}_{\lam_3}\right)^{\dagger}(n;\lam_0, \mu_0) \left[\figbox{0.66}{unassoc_app}\right]_{\calF}
\ee
 As the associator corresponds to a change of basis in a space of intertwiners, it inverse can be found by taking its conjugate with respect to an inner product. In other words, abbreviating all parameters except $\lam_3$, $\lam_0$, and $n$ in \eqref{dissrel} and denoting the diagram on the RHS and LHS by $v_{\lam_3, \lam_0}$ and $w_{\lam_3, n}$, respectively,
 \be
 F_{\lam_3}\cdot v_{\lam_3, \lam_0}=v_{\lam_3, \lam_0}=\sum_n F_{\lam_3}(n; \lam_0)w_{\lam_3, n}
 \ee
 so assuming some Hermitian inner product on the space of maps $\calF_{\lam_3}^{\mu_3} \to \calF^{\mu_4}_{\lam_4} \otimes \calF^{\mp}_{\De_1}\otimes\calF^{\mp}_{\De_2}$,
 \be
 \bcorr{v_{\lam'_3, \lam'_0}, v_{\lam_3, \lam_0}}=\bcorr{F_{\lam'_3}v_{\lam'_3, \lam'_0}, F_{\lam_3}v_{\lam_3, \lam_0}}=\bcorr{v_{\lam'_3, \lam'_0}, F^{\dagger}_{\lam_3}F_{\lam_3}v_{\lam_3, \lam_0}}.
 \ee
 There is a natural Hermitian inner product to use, using which for example
 \begin{align} \label{innprodapp}
 \bcorr{v_{\lam'_3, \lam'_0}, v_{\lam_3, \lam_0}}&=\left[\figbox{0.66}{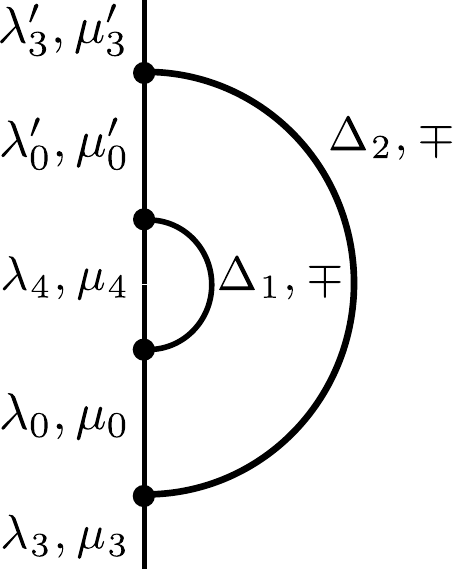}\right]_{\calF}=\de(E_3-E'_3)\de(E_0-E'_0)N_{\lam_3, \mu_3}^{\lam_0, \De_2}N_{\lam_0, \mu_0}^{\lam_4, \De_1},\nn
  \bcorr{w_{\lam'_3, n'_0}, w_{\lam_3, n}}&=\left[\figbox{0.66}{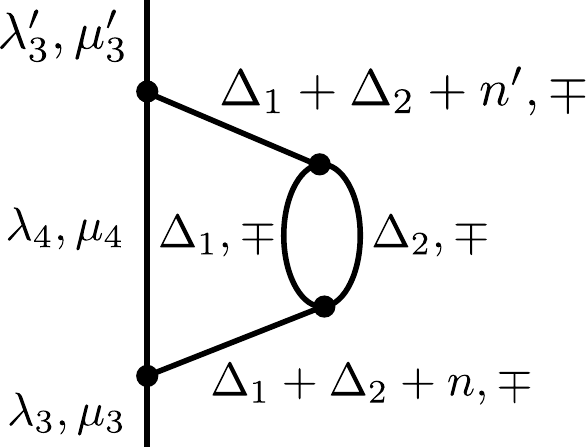}\right]_{\calF}=\de(E_3-E'_3)\de_{n, n'}d_n N_{\lam_3, \mu_3}^{\lam_4, \De_1+\De_2+n}.
 \end{align}
 Then from the definition of the conjugate
 \be
 \bcorr{v_{\lam'_3, \lam'_0},F^{\dagger}_{\lam_3}w_{\lam_3, n}}=\bcorr{F_{\lam'_3}v_{\lam'_3, \lam'_0}, w_{\lam_3, n}}=\bcorr{w_{\lam_3, n}, F_{\lam'_3}v_{\lam'_3, \lam'_0}}^*
 \ee
we find that matrix elements of \eqref{dissrel} are given by
\be
\left(F^{\lam_4, \De_1, \De_2, \mp}_{\lam_3}\right)^{\dagger}(n;\lam_0, \mu_0)=F^{\lam_4, \De_1, \De_2, \mp}_{\lam_3}(\lam_0, \mu_0; n)^* {d_n N_{\lam_3, \mu_3}^{\lam_4, \De} \ov N_{\lam_3, \mu_3}^{\lam_0, \De_2} N_{\lam_0, \mu_0}^{\lam_4, \De_1}}, \qquad \De=\De_1+\De_2+n.
\ee

The final step is to dress the diagrams in \eqref{assrel} and \eqref{dissrel} with relative coefficients in \eqref{bratio} and \eqref{mratio}, in steps analogous to \eqref{ratiouse}. Then we have for coefficients in \eqref{ass} and \eqref{diss}
\begin{align}
F^{E_4, \De_1, \De_2}_{E_3}(E_0; n)&=F^{\lam_4, \De_1, \De_2, \mp}_{\lam_3}(\lam_0, \mu_0; n) { C_{\lam_3, \mu_3}^{\lam_0, \De_2, \mp}C_{\lam_0, \mu_0}^{\lam_4, \De_1, \mp} \ov C_{\lam_3, \mu_3}^{\lam_4, \De, \mp}C_{n}^{\De_1, \De_2}},\\
\left(F^{E_4, \De_1, \De_2}_{E_3}\right)^{\dagger}(n; E_0)&=F_{E_3}^{E_4, \De_1, \De_2}(E_0; n)^*{d_n N_{\lam_3, \mu_3}^{\lam_4, \De} \ov N_{\lam_3, \mu_3}^{\lam_0, \De_2} N_{\lam_0, \mu_0}^{\lam_4, \De_1}}\left|{  C_{\lam_3, \mu_3}^{\lam_4, \De, \mp}C_{n}^{\De_1, \De_2} \ov C_{\lam_3, \mu_3}^{\lam_0, \De_2, \mp}C_{\lam_0, \mu_0}^{\lam_4, \De_1, \mp} }\right|^2
\end{align}
which simplify to the results in \eqref{diag4}, \eqref{diag5}, and \eqref{Tamp}. These coefficients satisfy the orthogonality relation
\begin{align}
&\int dE_0\,  \left(F_{E_3}^{E_4, \De_1, \De_2}\right)^{\dagger}(n; E_0)\, F_{E_3}^{E_4, \De_1, \De_2}(E_0; n')=\nn
&\hspace{100pt}{\Ga(2 \De_1 + 2 \De_2 + n-1) \ov n!}{(2 \De-1) \ov \Ga(2 \De_1 +n)\Ga(2 \De_2 +n)\Ga(\De \pm i s_3 \pm i s_4)}\de_{n, n'}
\end{align}
which follows from the orthogonality relation for Wilson polynomials first derived in \cite{Wil80}.

\bibliography{Sch_corr}

\providecommand{\href}[2]{#2}\begingroup\raggedright\begin{thebibliography}{10}

\bibitem{Ja85}
R.~Jackiw, \emph{Lower dimensional gravity},
  \href{https://doi.org/10.1016/0550-3213(85)90448-1}{\emph{Nuclear Physics B}
  {\bfseries 252} (1985) 343 -- 356}.

\bibitem{Te83}
C.~Teitelboim, \emph{{Gravitation and Hamiltonian structure in two space-time
  dimensions}}, \href{https://doi.org/10.1016/0370-2693(83)90012-6}{\emph{Phys.
  Lett.} {\bfseries 126B} (1983) 41--45}.

\bibitem{AlPo14}
A.~Almheiri and J.~Polchinski, \emph{{Models of AdS$_{2}$ backreaction and
  holography}}, \href{https://doi.org/10.1007/JHEP11(2015)014}{\emph{JHEP}
  {\bfseries 11} (2015) 014},
  [\href{https://arxiv.org/abs/1402.6334}{{\ttfamily 1402.6334}}].

\bibitem{SaYe93}
S.~Sachdev and J.~Ye, \emph{{Gapless spin fluid ground state in a random,
  quantum Heisenberg magnet}},
  \href{https://doi.org/10.1103/PhysRevLett.70.3339}{\emph{Phys. Rev. Lett.}
  {\bfseries 70} (1993) 3339},
  [\href{https://arxiv.org/abs/cond-mat/9212030}{{\ttfamily
  cond-mat/9212030}}].

\bibitem{Kit.KITP}
A.~Kitaev, ``A simple model of quantum holography.'' Talks at KITP
  \url{http://online.kitp.ucsb.edu/online/entangled15/kitaev/} and
  \url{http://online.kitp.ucsb.edu/online/entangled15/kitaev2/}, April and May,
  2015.

\bibitem{SoftMode}
A.~Kitaev and S.~J. Suh, \emph{{The soft mode in the Sachdev-Ye-Kitaev model
  and its gravity dual}},
  \href{https://doi.org/10.1007/JHEP05(2018)183}{\emph{JHEP} {\bfseries 05}
  (2018) 183}, [\href{https://arxiv.org/abs/1711.08467}{{\ttfamily
  1711.08467}}].

\bibitem{KiSuh18}
A.~Kitaev and S.~J. Suh, \emph{{Statistical mechanics of a two-dimensional
  black hole}}, \href{https://doi.org/10.1007/JHEP05(2019)198}{\emph{JHEP}
  {\bfseries 05} (2019) 198},
  [\href{https://arxiv.org/abs/1808.07032}{{\ttfamily 1808.07032}}].

\bibitem{MeTuVe17}
T.~G. Mertens, G.~J. Turiaci and H.~L. Verlinde, \emph{{Solving the Schwarzian
  via the Conformal Bootstrap}},
  \href{https://doi.org/10.1007/JHEP08(2017)136}{\emph{JHEP} {\bfseries 08}
  (2017) 136}, [\href{https://arxiv.org/abs/1705.08408}{{\ttfamily
  1705.08408}}].

\bibitem{Lam18}
H.~T. Lam, T.~G. Mertens, G.~J. Turiaci and H.~Verlinde, \emph{Shockwave
  s-matrix from schwarzian quantum mechanics},
  \href{https://doi.org/10.1007/jhep11(2018)182}{\emph{Journal of High Energy
  Physics} {\bfseries 2018} (Nov, 2018) }.

\bibitem{SL2R}
A.~Kitaev, \emph{Notes on {$\widetilde{\mathrm{SL}}(2,\mathbb{R})$}
  representations},  \href{https://arxiv.org/abs/1711.08169}{{\ttfamily
  1711.08169}}.

\bibitem{Gro05}
W.~Groenevelt, \emph{Wilson function transforms related to racah coefficients},
   \href{https://arxiv.org/abs/math/0501511}{{\ttfamily math/0501511}}.

\bibitem{Wil80}
J.~A. Wilson, \emph{Some hypergeometric orthogonal polynomials},
  \href{https://doi.org/10.1137/0511064}{\emph{SIAM Journal on Mathematical
  Analysis} {\bfseries 11} (1980) 690--701},
  [\href{https://arxiv.org/abs/https://doi.org/10.1137/0511064}{{\ttfamily
  https://doi.org/10.1137/0511064}}].

\bibitem{Saad19}
P.~Saad, \emph{Late time correlation functions, baby universes, and eth in jt
  gravity},  2019.

\end{thebibliography}\endgroup
\bibliographystyle{JHEP}
 
\end{document}